\documentclass[a4paper]{jpconf}
\usepackage{graphicx}
\usepackage{html}
\usepackage{subcaption}
\begin{document}
\title{Designing Computing System Architecture and Models for the HL-LHC era}

\author{L Bauerdick$^1$, B Bockelman$^2$, P Elmer$^3$, S Gowdy$^1$, M
  Tadel$^4$ and F W\"urthwein$^4$}

\address{$^1$Fermilab, Batavia, IL 60510, USA\\
  $^2$Computer Science \& Engineering, University of Nebraska-Lincoln,
  Lincoln, NE 68588, USA\\
  $^3$Department of Physics, Princeton University, Princeton, NJ 08540, USA\\
  $^4$Department of Physics, UCSD, La Jolla, CA  92093, USA}

\ead{sgowdy+chep15@gmail.com}

\begin{abstract}
This paper describes a programme to study the computing model in CMS after 
the next long shutdown near the end of the decade.
\end{abstract}

\section{Introduction}

One of the recurring challenges for HEP computing in recent years has
been data management, access and organisation when using distributed
computing resources.  The computing model chosen by CMS for the LHC
startup used a distributed data management architecture~\cite{CMSCTDR}
which placed datasets statically at sites. Dataset replicas in
multiple sites were made manually as required, and jobs were sent to
sites where their input data could be read from site-local storage.
The wide area network (WAN) was underutilised as a resource, despite
being significantly more robust than originally imagined.

The reliability of all WLCG computer centres has greatly improved
through the experience gained during LHC Run 1. More sophisticated
data management and access models are thus possible. The use of
``opportunistic'' compute resources also becomes much easier, as they
can be used without requiring local data storage.  For LHC Run 2 CMS
is deploying additional technologies to monitor dataset popularity and
use PhEDEx~\cite{PHEDEX} to remove unnecessary dataset
replications. These evolutionary changes will allow effective data
management and access through Run 2.

\section{Data Management at the HL-LHC}
Planning is currently underway for a High Luminosity Large Hadron
Collider (HL-LHC)~\cite{HLLHC} with an objective of accumulating
${3000fb^{-1}}$ by 2030. Taking into account the upcoming increase
in energy for Run 2, and expectations for evolving pile-up and
trigger rate through Run 3 and HL-LHC, the data volume increase
over the next 15 years will be O($10^3$).

This may imply that HL-LHC (and perhaps Run 3) will
require larger, potentially non-evolutionary, changes to the
experiment's computing model, data management, access and organisation.
To that end, a number of ideas and research questions have arisen
in the community: Can the architectures and algorithms used for
caching and to reduce latencies in Content Delivery Networks be
applied to building such a system?  What can be learned from
commercial/general purpose cloud storage systems (e.g.\ Google
Drive, DropBox) to evolve the existing data federation into a
cost-effective, high performance global storage cloud for physics?
How can HEP best exploit a hierarchy of cache storage
from client side memory, through SSD's and disks to tape?  As the
importance of the WAN is increased, are there specific technologies
(beyond simple bandwidth increases) that can help?

\section{Computing Model Simulation}

To aid in deciding which type of model could produce the most
efficient use of resources a simulation has been developed. The first
results from that simulation under three different scenarios are
described in this paper.

\subsection{Description}

The \htmladdnormallink{simulation}{https://github.com/gowdy/sitesim}
is a event driven discrete simulation. The events are merely time
slices. In the results reported in this paper the time slices are
every 100 seconds.

\begin{figure}
  \includegraphics[trim=100 140 100 130, clip, width=\textwidth]{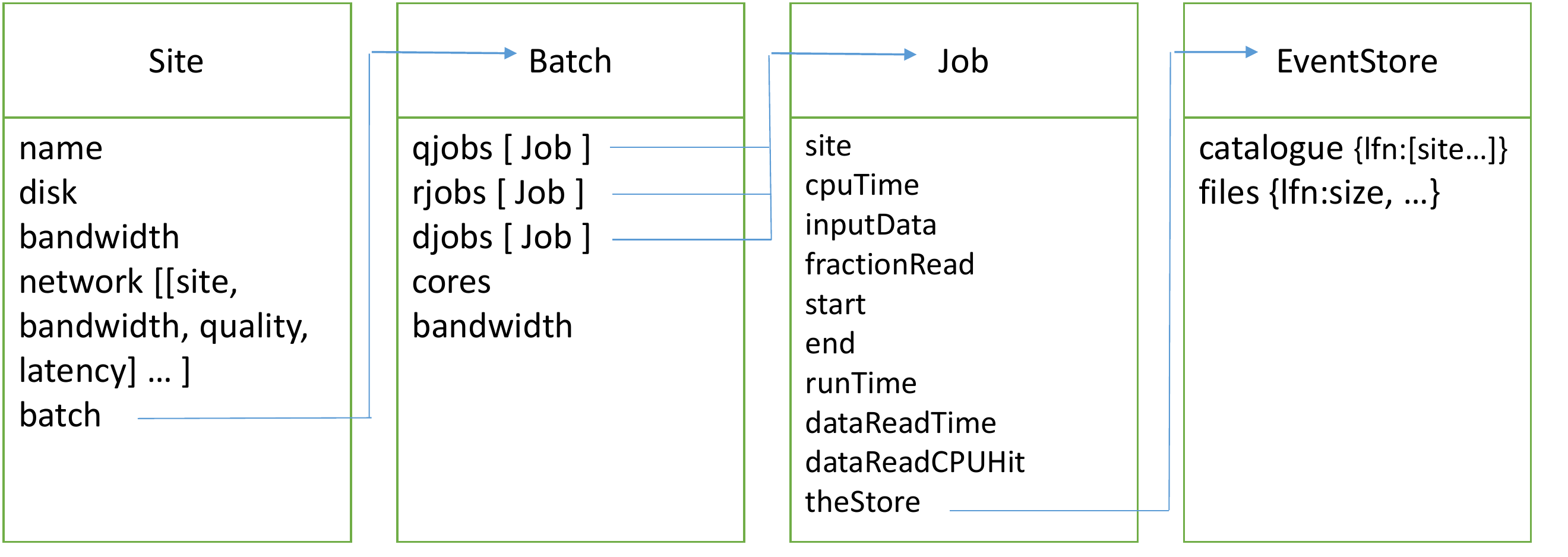}
  \caption{Software classes in the CMS Computing Model
    Simulation\label{fig:classDiag}}
\end{figure}

In the simulation each site is defined. A site also contains a batch
system, a disk storage system, and a set of network links to other
sites. The relationship between the software defined components is
shown in Figure \ref{fig:classDiag}.

The batch system has a set of cores for running jobs. It also
contains, but doesn't currently use, a internal site bandwidth which
could further constrain the speed of jobs running at that site. It
maintains a list of jobs for each state, queued, running or done.

The disk storage system is configured as an available resource. Files
can be stored locally and use up this space. There is no tertiary data
storage system defined.

The network links to other sites are defined with information on the
bandwidth of the link, the latency implicit in that link and the quality
of the link. This information is used to determine how fast data will
flow over the links.

In addition there is also an Event Store that is used to define
information about files used in the system. It contains the Logical
File Name (LFN) and the size of each file. It also has knowledge of
which sites have the files stored.

The information defined for a job is the CPU time required to carry it
out, the site it will run at, which LFNs it will read and what
fraction of that data it will read (currently defined to be 100\%). It
also remembers the original wall clock time required to run the job.

\subsection{Information Sources}

The simulation uses current system information to setup a complex and
realistic experiment wide computing system. Site information can be
extracted from the CMS
\htmladdnormallink{SiteDB}{http://cmsweb.cern.ch/sitedb/}
service. This provides information on the resources pledged to the
experiment for disk space and CPU power (defined in HEPSpec06). There
is an automated tool to extract this information (which was the most
recent 2014 pledges) and format it as input for the simulation. In
addition, when only considering the CMS infrastructure in the US these
numbers were extracted from
\htmladdnormallink{REBUS}{http://wlcg-rebus.cern.ch/apps/pledges/resources/}.

Once the sites are setup the links between them get setup then using
information extracted by a script from
\htmladdnormallink{PhEDEx}{http://cmsweb.cern.ch/phedex/}. This
provides a list of links between the sites. Each of those links can
have a quality associated with it. This provides information on how
often file transfers need to be retried. In addition information on
the actual transfer rate is available. However, this information is
only available as an aggregated number. It used as an indication of
the bandwidth available on the link. This number is constrained to be
between 1GB/s and 10GB/s. The other information associated with a link
is the latency of the link. Currently this number is estimated based
on the distance between sites (see Table \ref{tab:latency} for the
values used). In a future update to the simulation this should be an
relatively easy number to measure.

\begin{table}
  \begin{center}
    \begin{footnotesize}
      \begin{tabular}{|l|rrrrrrrrr|}
        \hline
        Site & Purdue & UCSD & Nebraska & Wisconsin & Vanderbilt & Caltech & Florida & MIT & FNAL \\
        \hline
        Purdue & 0 & 100 & 100 & 60 & 40 & 100 & 40 & 40 & 70 \\
        UCSD & 100 & 0 & 70 & 100 & 100 & 20 & 100 & 100 & 100 \\
        Nebraska & 70 & 60 & 0 & 40 & 70 & 40 & 70 & 70 & 40 \\
        Wisconsin & 40 & 70 & 40 & 0 & 60 & 100 & 70 & 40 & 20 \\
        Vanderbilt & 40 & 100 & 70 & 70 & 0 & 100 & 40 & 20 & 60 \\
        Caltech & 100 & 20 & 60 & 100 & 100 & 0 & 100 & 100 & 100 \\
        Florida & 40 & 100 & 70 & 60 & 40 & 100 & 0 & 60 & 70 \\
        MIT & 40 & 100 & 100 & 70 & 40 & 100 & 40 & 0 & 70 \\
        FNAL & 40 & 100 & 40 & 20 & 70 & 100 & 70 & 60 & 0 \\
        \hline
      \end{tabular}
      \caption{Link latency (ms) from (horizontal) site to (vertical) site\label{tab:latency}}
    \end{footnotesize}
  \end{center}
\end{table}

Once sites and the links between them are setup the file size and
location information are loaded. These have also been extracted from
PhEDEx. The list of files required is found from the list of jobs to
be run. When considering only US locations, files that are needed but
not present in the US are artificially given a location at Fermilab,
the US Tier-1 site.

\subsection{Simulation Parameters}

There are a few distributions used as parameters of the
simulation. These are also read from flat files but the information is
from different sources.

The first of these is the drop in CPU efficiency seen when running
jobs that access data from a remote storage element. For example a job
reading data at Fermilab while running at UCSD could drop from a 95\%
efficiency to 75\% CPU efficiency. The numbers used currently are
merely an estimate till more accurate information can be
gathered. These are tabulated in Table \ref{tab:cpuHIT}.

\begin{table}
  \begin{center}
    \begin{tabular}{|l|r|}
      \hline
      Latency (ms) & CPU Efficiency Penalty (\%) \\
      \hline
      0 (ie same site) & 0 \\
      $>=$1ms & 5 \\
      $>=$50ms & 20 \\
      \hline
    \end{tabular}
    \caption{CPU Efficiency Penalty as a function of link latency\label{tab:cpuHIT}}
  \end{center}
\end{table}

Another parameter is the maximum single file transfer rate of a given
link. This is again based on the link latency. The standard values are
shown in Table \ref{tab:linkLatency}.

\begin{table}
  \begin{center}
    \begin{tabular}{|l|r|}
      \hline
      Latency (ms) &  Maximum Single File Transfer Speed (MB/s) \\
      \hline
      0 (ie same site) & 10000 \\
      $>=$1ms & 1000 \\
      $>=$50ms & 100 \\
      $>=$100ms & 50 \\
      \hline
    \end{tabular}
    \caption{CPU Efficiency Penalty as a function of link
      latency\label{tab:linkLatency}}
  \end{center}
\end{table}

The last set of parameters used is to Monte-Carlo the CPU efficiency
of jobs. A set of jobs run in September of last year is used as the
sample set. A distribution is derived from them, which is also binned
by CPU time as it is observed that shorter jobs can have a much worse
CPU efficiency than longer jobs. There are 100 CPU efficiency bins and
10 CPU time bins in the distribution.

\section{Scenarios}

\begin{figure}
  \includegraphics[trim=80 90 80 90, clip, width=\textwidth]{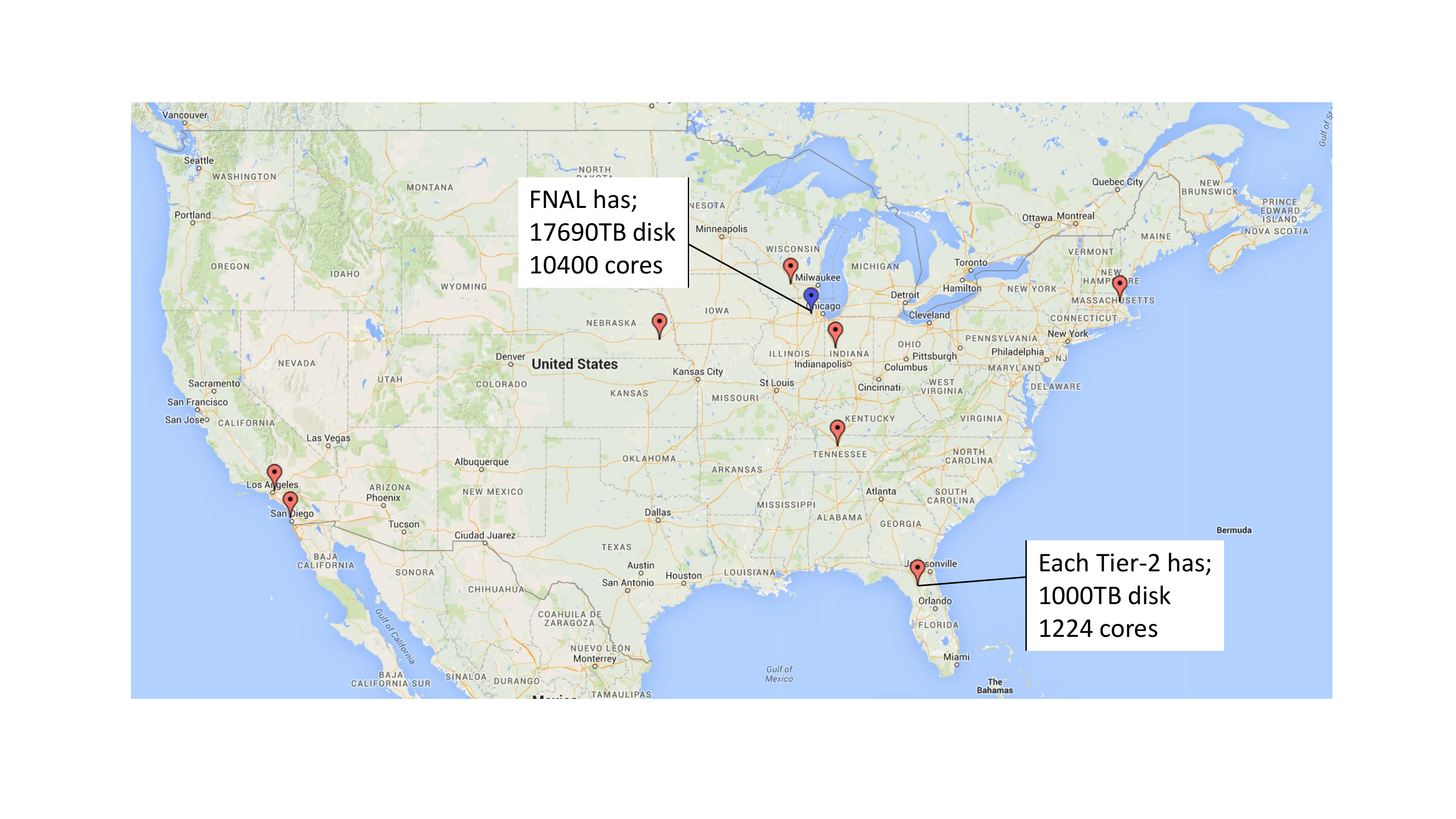}
  \caption{Sites used in the simulation\label{fig:map}}
\end{figure}

In each case the CMS computing system in the US is used. This consists
of one Tier-1 site (FNAL) and eight Tier-2 sites. Figure \ref{fig:map}
shows the location of these sites together with the resources
available there today according to REBUS.

\subsection{Data preplaced at sites\label{sec:today}}

In this scenario data is already mostly preplaced at the site where
the job will execute. This is the situation today where bulk data
transfers are done using PhEDEx and in the vast majority of cases data
is already at a site before the job starts running. In the simulation
there are some jobs which do access remote data. In addition there
were some jobs where the data read wasn't present anywhere in the
US. In this latter case a copy of the data was placed at FNAL to allow
them to run.

\subsection{Data Replication}

With this scenario data is replicated to sites as the job starts. This
introduces a small inefficiency, depending on how long the job would
run. In this case all data is placed at FNAL and every job run outside
FNAL would copy the data it requires, except in the case where an
earlier job had already used the data in question. Copies are cached
at sites in this scenario.

\subsection{Remote Reading}

In this scenario all data is located at FNAL and nowhere else. Each
job must read over the network to access its data. In this scenario
there is no disk space required at the Tier-2s.

\section{Execution}

When execution begins all jobs are read in. For the results presented
56949 jobs were extracted from the dashboard, as run during a week in
February 2015. To provide a more even distribution of jobs across the
infrastructure each job was duplicated and for the Tier-2 based jobs,
ran at another site. For the first scenario described in Section
\ref{sec:today} the data was also duplicated with the job. The jobs
are added to the queue at the sites they are to be run at.

Each site is then polled to start any jobs it has capacity to
start. Then data transfers are constrained for those jobs, this is
needed to determine the run time of the job. Jobs are also checked to
see if they've finished since the last poll. This may allow some
transfers to complete sooner, and hence the jobs that depend on them.

Each job has the CPU efficiency Monte-Carlo run for it to determine a
first order wall clock time. This can be further extended if data
movement is required for the job. A site to site transfer would add a
fixed amount of time. A remote read will add a CPU Efficiency penalty
while that file is being read. The CPU time of the job is shared
between the files based on the size of the file. The simulation takes
account of bad quality links while transferring files and has a
mechanism to generate retries, and will eventually give up if too many
retries occur. It would then locate another copy of the file to
use. Once the job does complete there is a total wall clock time
recorded for the job.

\section{Results}

\subsection{Total Wall Clock Time}

We can use the total wall clock time required to run all the jobs in
each of the scenarios to make a comparison. Table \ref{tab:wallClock}
shows these total wall clock time for the three scenarios. It also
shows the results if we vary the simulation parameters by a factor of
two in each direction.

\begin{table}
  \begin{center}
    \begin{scriptsize}
      \begin{tabular}{|l|rrr|rrr|rrr|}
        \hline
        & \multicolumn{3}{|c|}{Half CPU Hit} & \multicolumn{3}{|c|}{Normal CPU
          Hit} & \multicolumn{3}{|c|}{Double CPU Hit} \\
        \cline{2-10}
        & Preplaced & Copy & Remote & Preplaced & Copy & Remote
        & Preplaced & Copy & Remote \\
        \hline
        Half Max Speed & 2.77 & 3.32 & 3.78 & 2.77 & 3.32 & 3.94 & 2.77
        & 3.32 & 4.25 \\
        Normal Max Speed & 2.77 & 3.32 & 3.78 & 2.77 & 3.32 & 3.94 & 2.77
        & 3.32 & 4.25 \\
        Double Max Speed & 2.77 & 3.32 & 3.78 & 2.77 & 3.32 & 3.94 & 2.77
        & 3.32 & 4.25 \\
        \hline
      \end{tabular}
      \caption{Total wall clock time of all jobs in billions of seconds\label{tab:wallClock}}
    \end{scriptsize}
  \end{center}
\end{table}

There is a very small difference in the Transfer File time with the
change in maximum single file transfer speed. This isn't evident as it
is only apparent beyond the first three significant figures.

You can see that in each set of parameters that not preplacing the
data costs almost 20\% more wall clock time to be used. The penalty
for doing remote reads varies from 36\% up to 53\%, depending on the
actual set of remote read parameters used. The simulation is more
sensitive to this parameter.

\subsection{CPU Efficiency}

\begin{figure}
  \begin{center}
    \hspace{2.7cm}
    \begin{subfigure}{0.3\textwidth}
      \includegraphics[width=\textwidth]{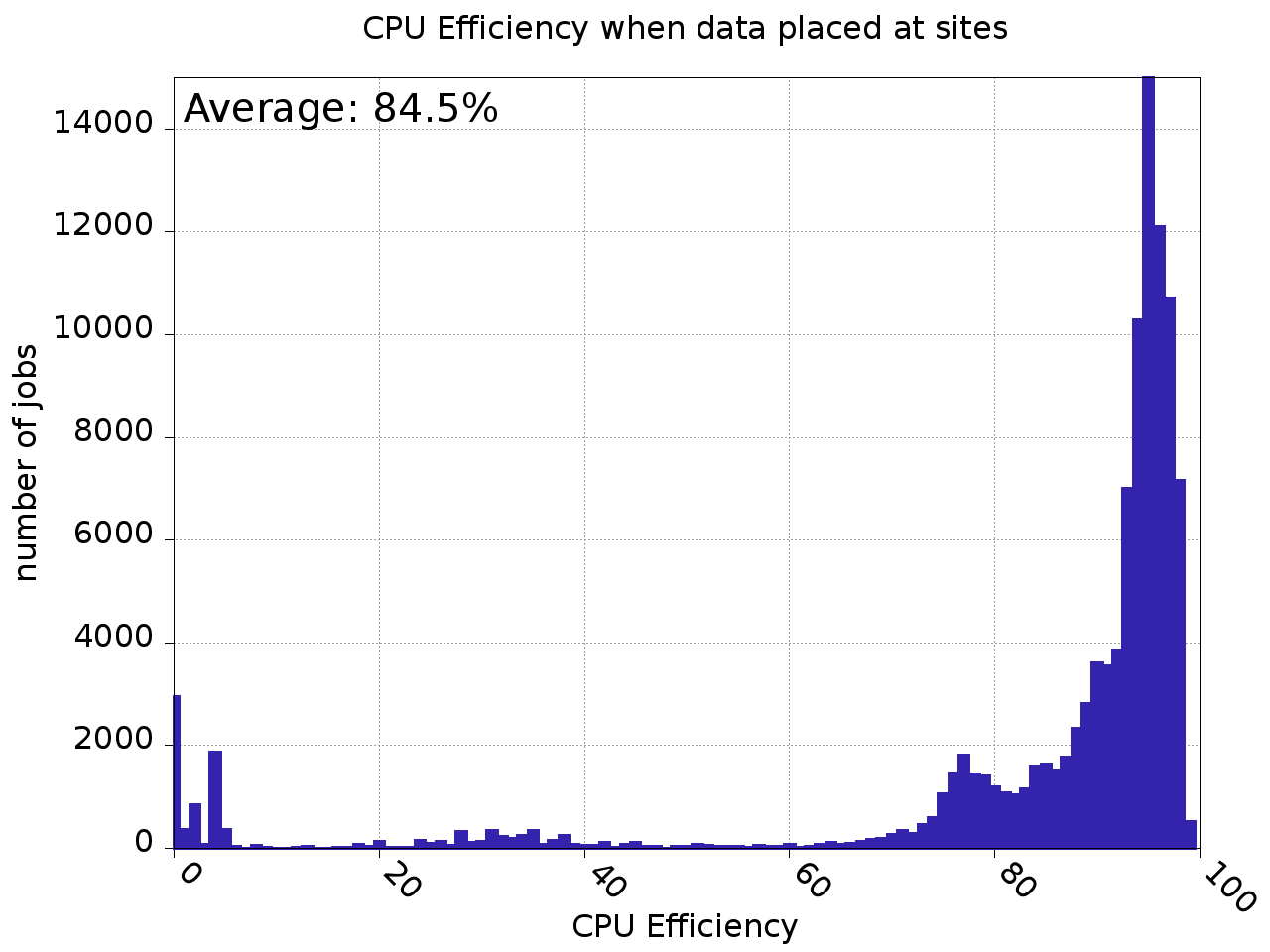}
      \caption{Preplaced Data\label{fig:cpuEffToday}}
    \end{subfigure}
    \newline
    \begin{subfigure}{0.3\textwidth}
      \includegraphics[width=\textwidth]{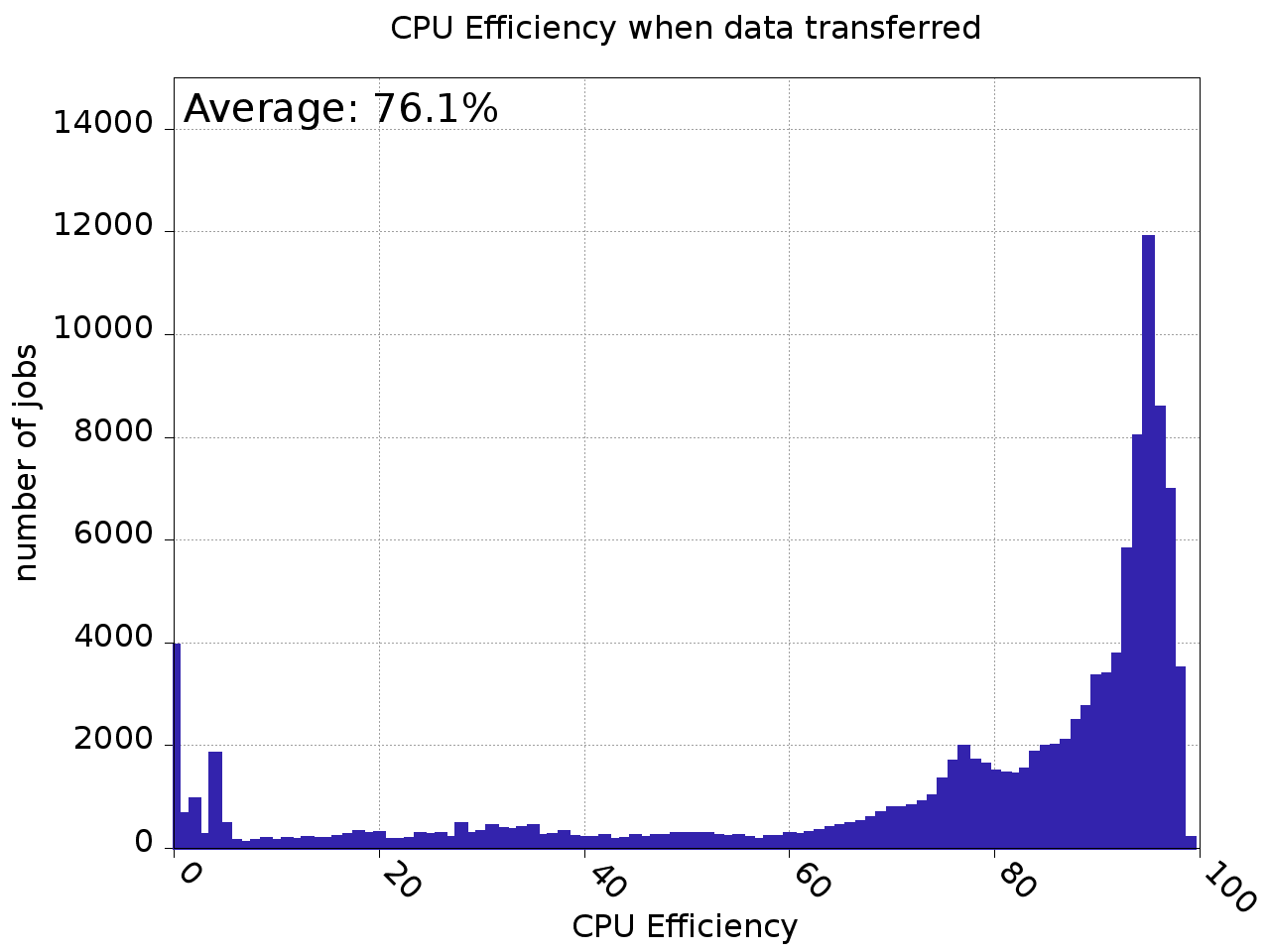}
      \caption{Copy: Half Rate\label{fig:cpuEffCopyH}}
    \end{subfigure}
    \begin{subfigure}{0.3\textwidth}
      \includegraphics[width=\textwidth]{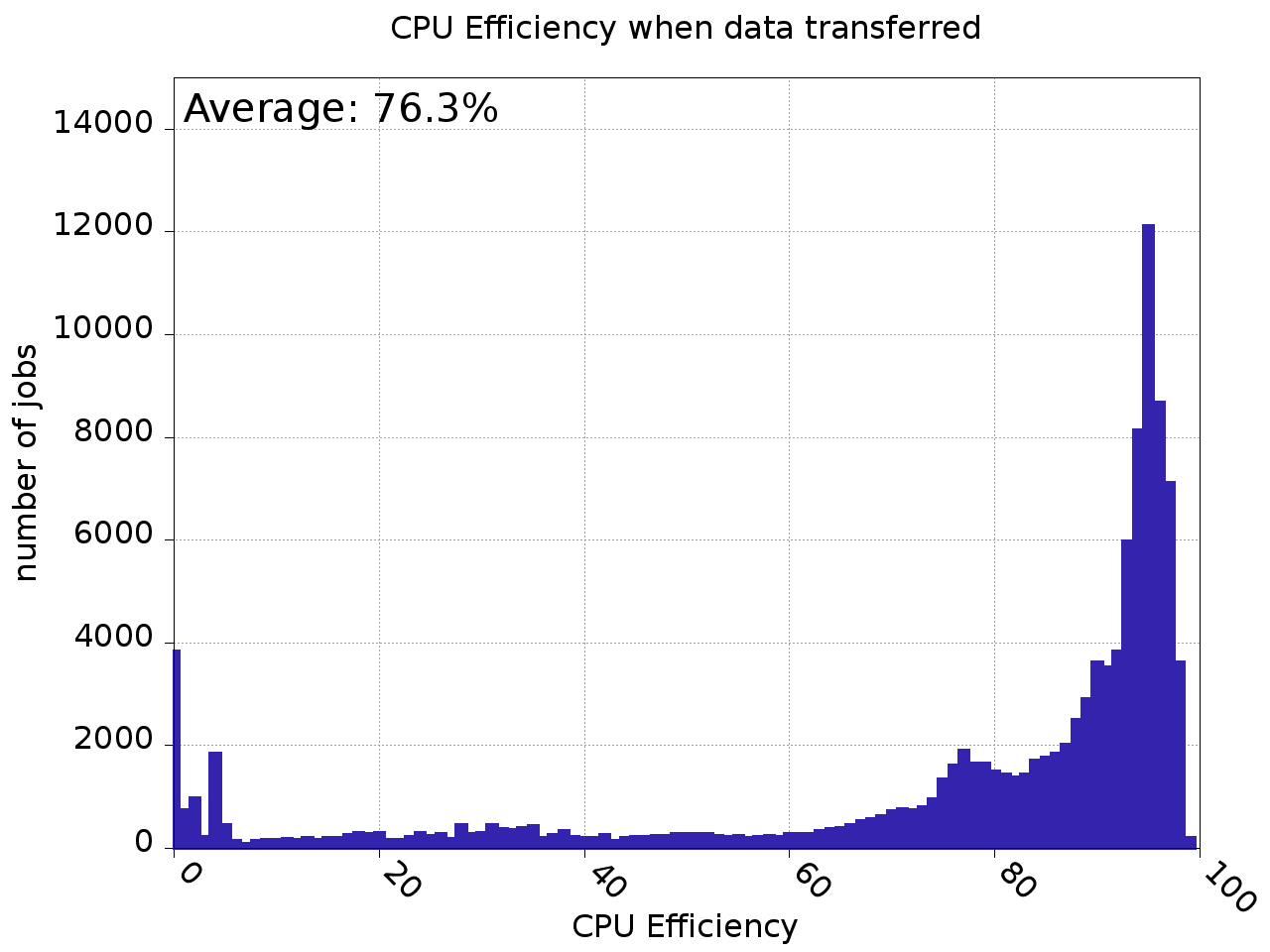}
      \caption{Copy: Normal Rate\label{fig:cpuEffCopyN}}
    \end{subfigure}
    \begin{subfigure}{0.3\textwidth}
      \includegraphics[width=\textwidth]{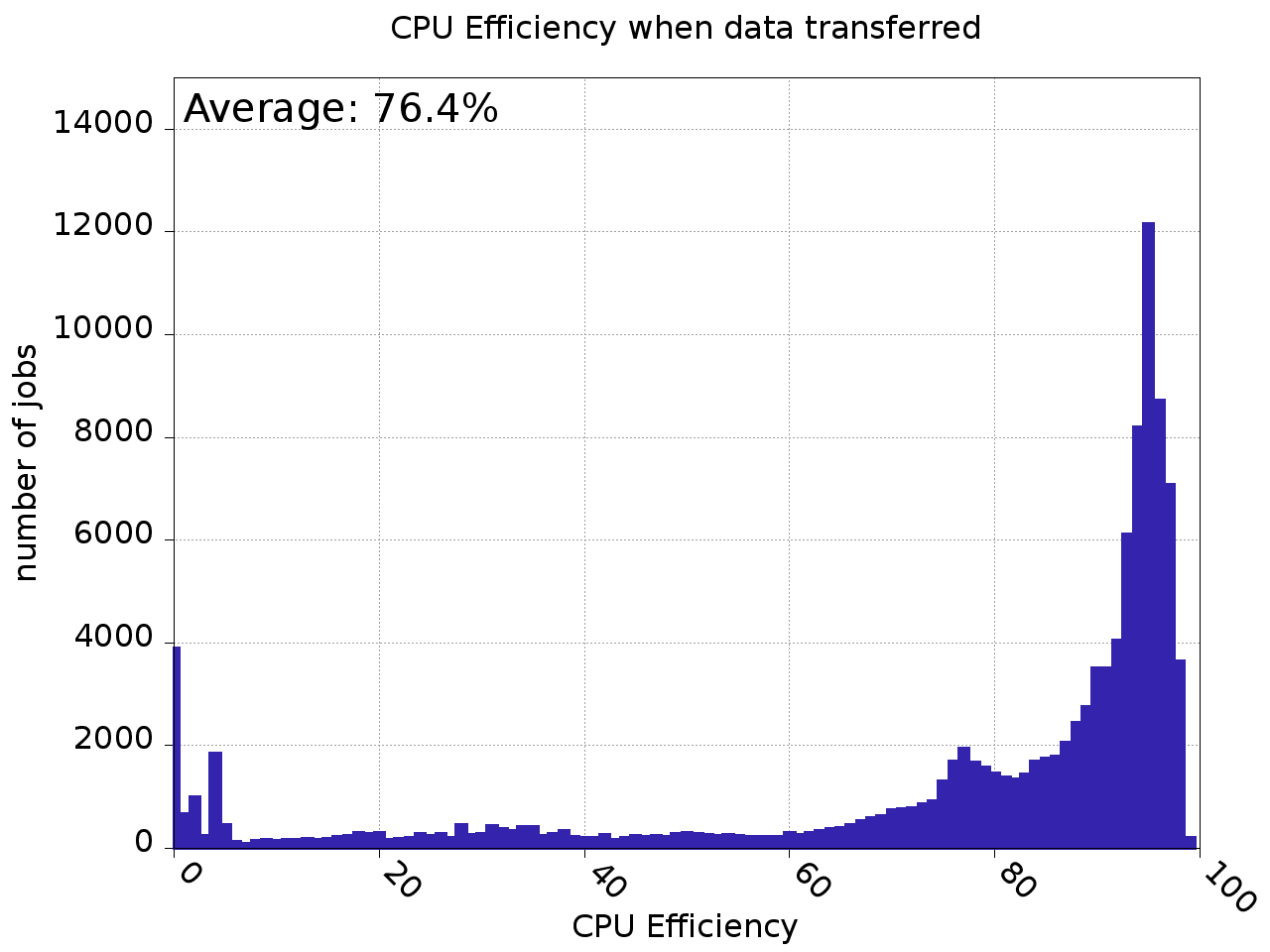}
      \caption{Copy: Double Rate\label{fig:cpuEffCopyD}}
    \end{subfigure}
    \begin{subfigure}{0.3\textwidth}
      \includegraphics[width=\textwidth]{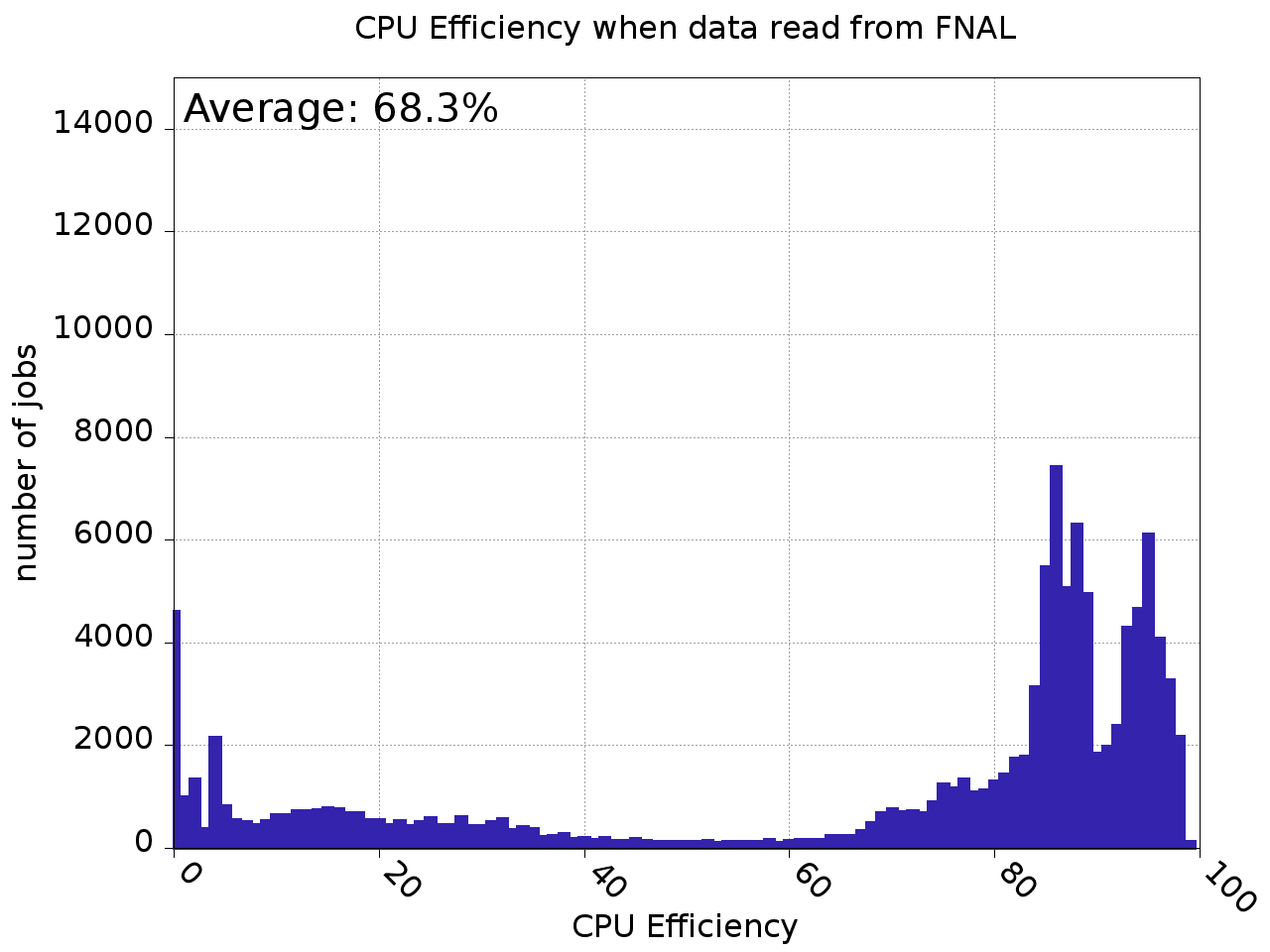}
      \caption{Remote: Half Penalty\label{fig:cpuEffRemoteH}}
    \end{subfigure}
    \begin{subfigure}{0.3\textwidth}
      \includegraphics[width=\textwidth]{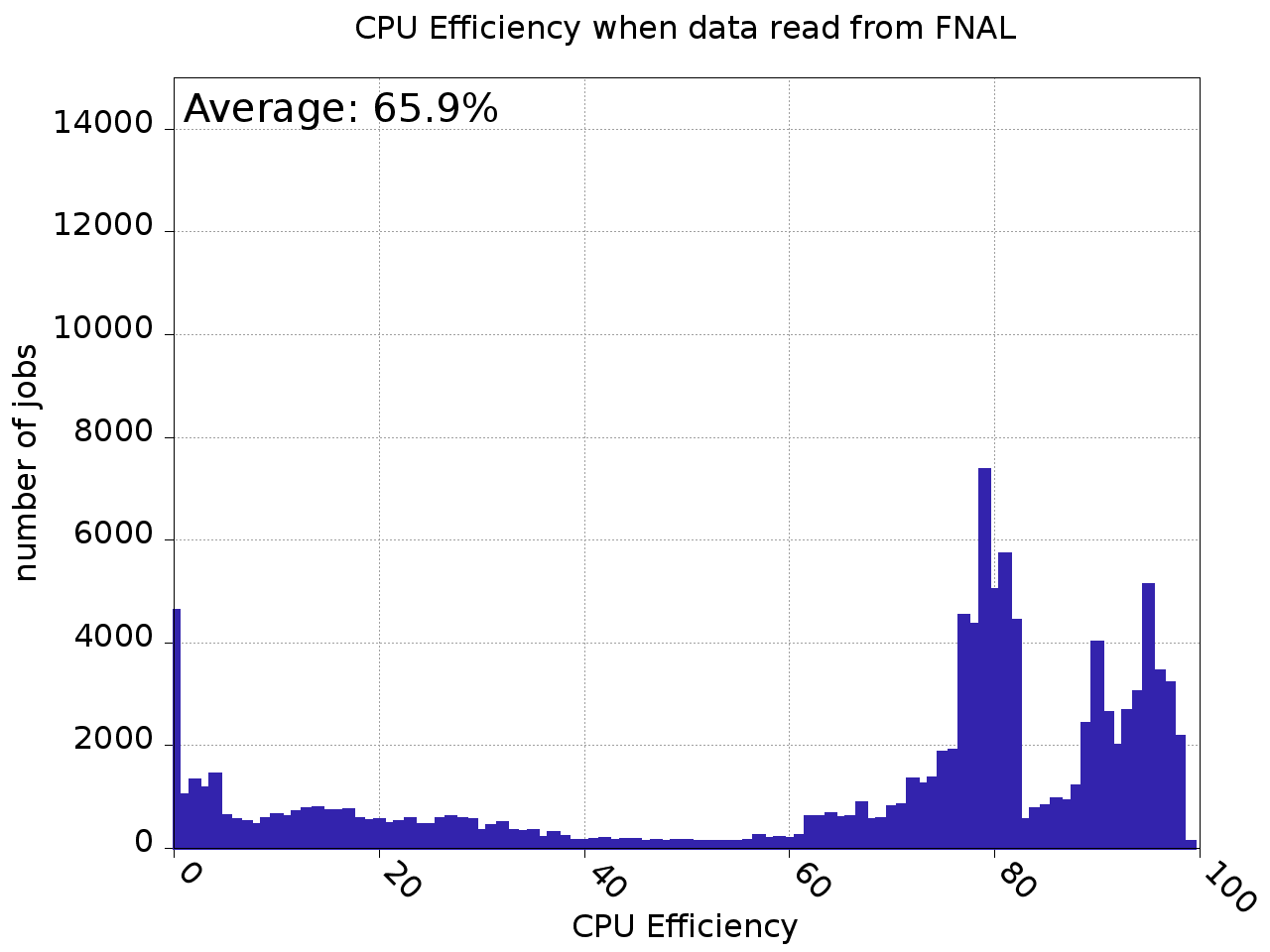}
      \caption{Remote: Normal Penalty\label{fig:cpuEffRemoteN}}
    \end{subfigure}
    \begin{subfigure}{0.3\textwidth}
      \includegraphics[width=\textwidth]{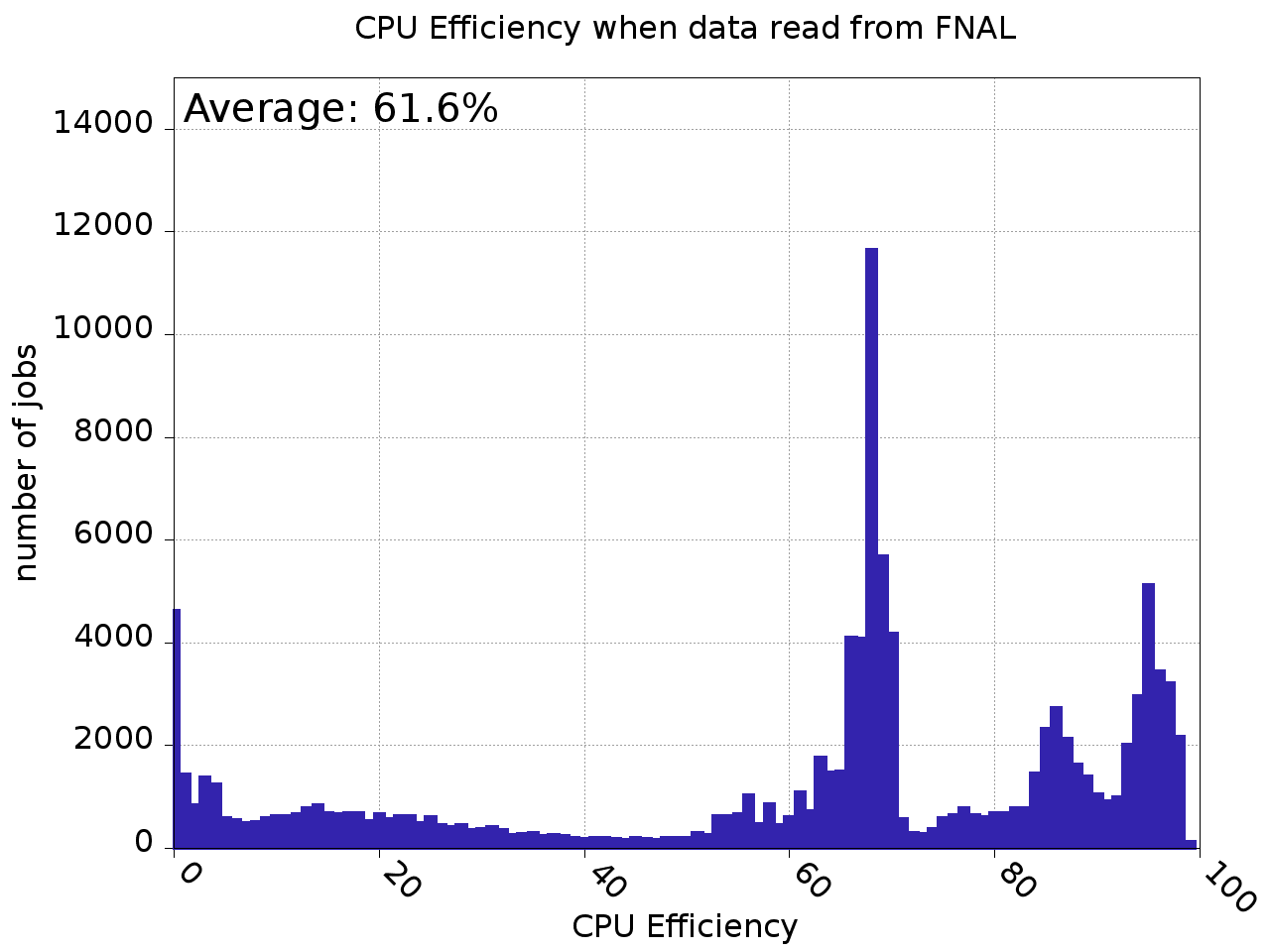}
      \caption{Remote: Double Penalty\label{fig:cpuEffRemoteD}}
    \end{subfigure}
    \caption{Distribution of CPU Efficiency across all
      jobs\label{fig:cpuEff}}
  \end{center}
\end{figure}

The latencies and penalties that jobs are exposed to cause their CPU
efficiency to drop. In the different scenarios we can see a slightly
different behaviour. We can see various CPU Efficiency distributions in
Figure \ref{fig:cpuEff}.

The distribution in Figure \ref{fig:cpuEffToday} shows the case where
data is preplaced at sites. There is very little change if the remote
read penalty or the maximum file transfer speed is varied. The average
CPU efficiency in this case is 84.5\%.

In Figures \ref{fig:cpuEffCopyH}-\ref{fig:cpuEffCopyD} you can see the
effect on the CPU efficiency when the maximum file transfer rate is
first halved, normal and then doubled. The effect is small, and the
average number varies from 76.1\% up to 76.4\%.

In the case of reading the data remotely there is a much larger
variation. This can be seen in Figures
\ref{fig:cpuEffRemoteH}-\ref{fig:cpuEffRemoteD}. Here the average
varies from 68.3\% down to 61.6\%. You can also see the artificial
nature of the imposed penalty producing a two peak structure. As the
penalty increases it moves the set of jobs with a lower penalty out of
those jobs run at FNAL, which have no penalty, producing three peaks.

\subsection{Job Queues During Simulation}

\begin{figure}
  \centering
  \begin{subfigure}{0.3\textwidth}
    \includegraphics[width=\textwidth]{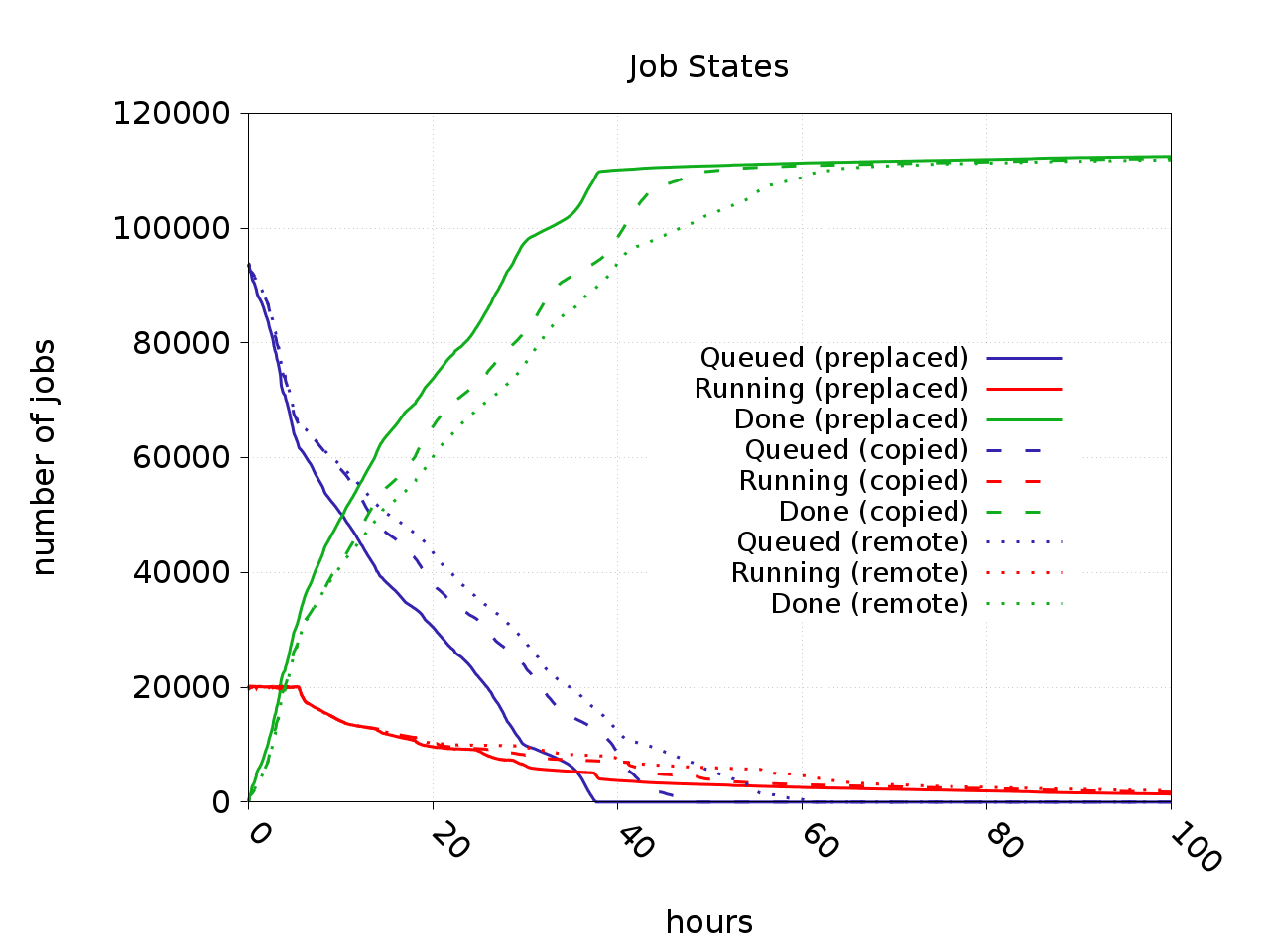}
    \caption{Half CPU/Half Tran}
  \end{subfigure}
  \begin{subfigure}{0.3\textwidth}
    \includegraphics[width=\textwidth]{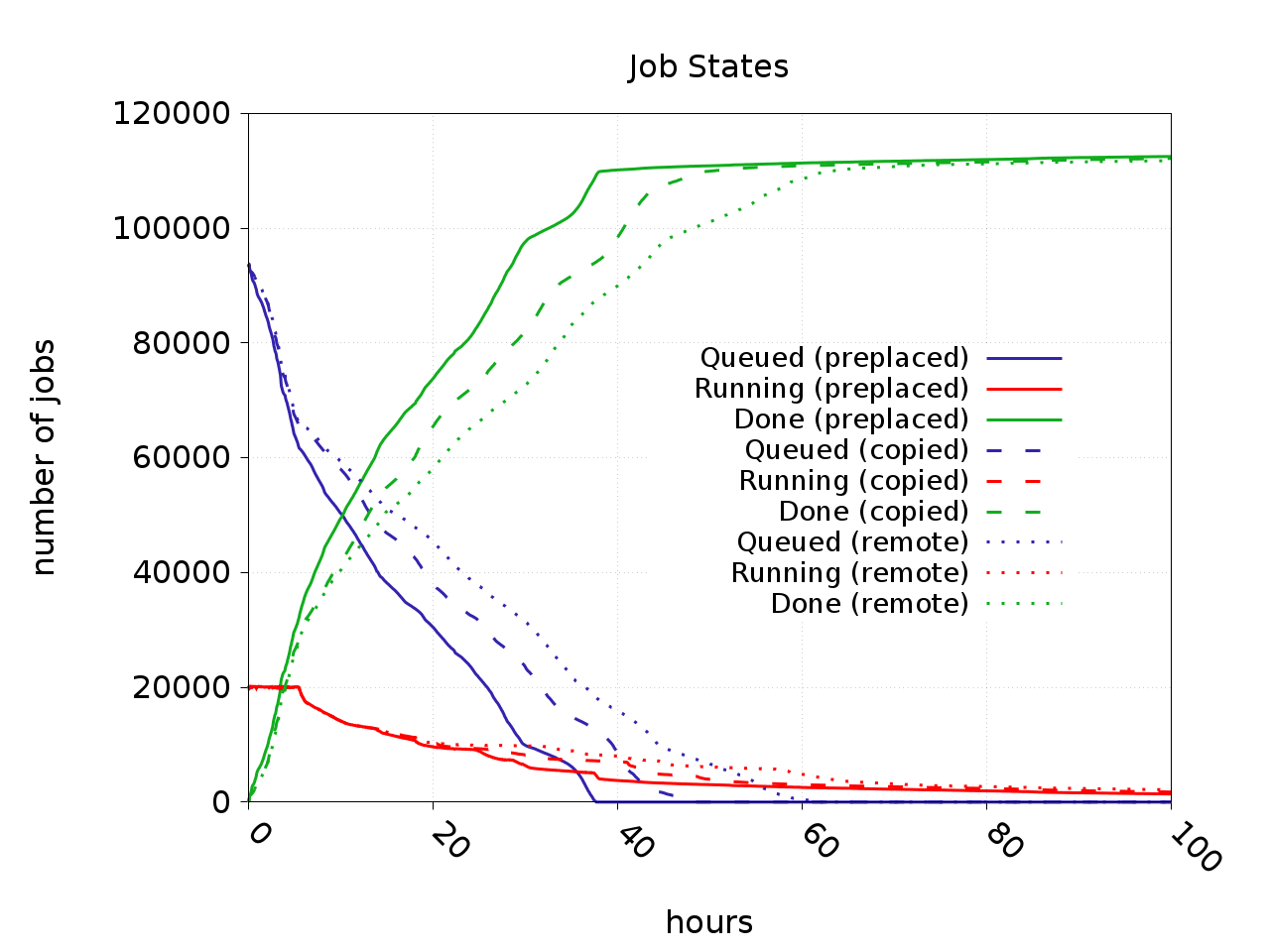}
    \caption{Normal CPU/Half Tran}
  \end{subfigure}
  \begin{subfigure}{0.3\textwidth}
    \includegraphics[width=\textwidth]{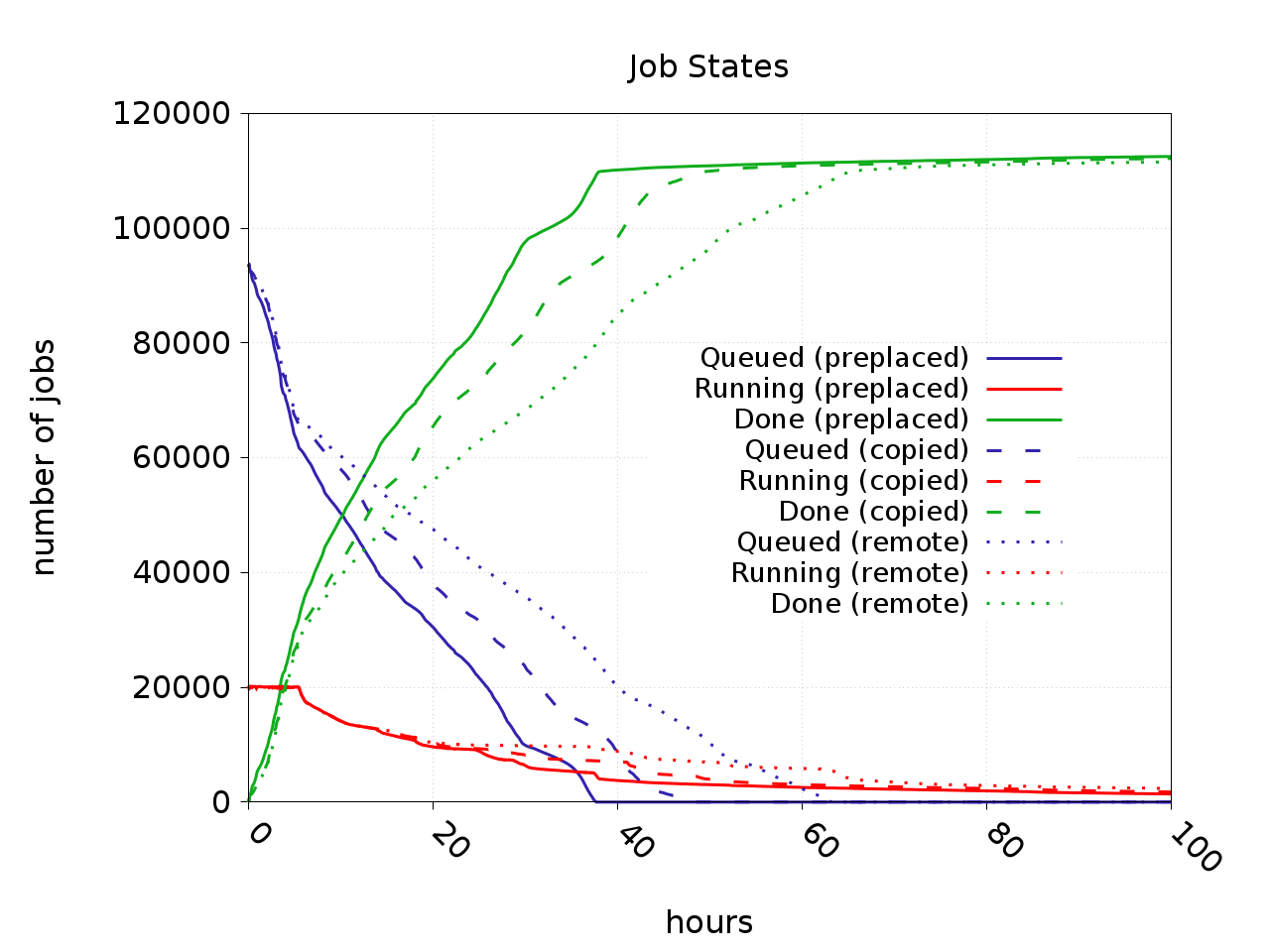}
    \caption{Double CPU/Half Tran}
  \end{subfigure}
  \begin{subfigure}{0.3\textwidth}
    \includegraphics[width=\textwidth]{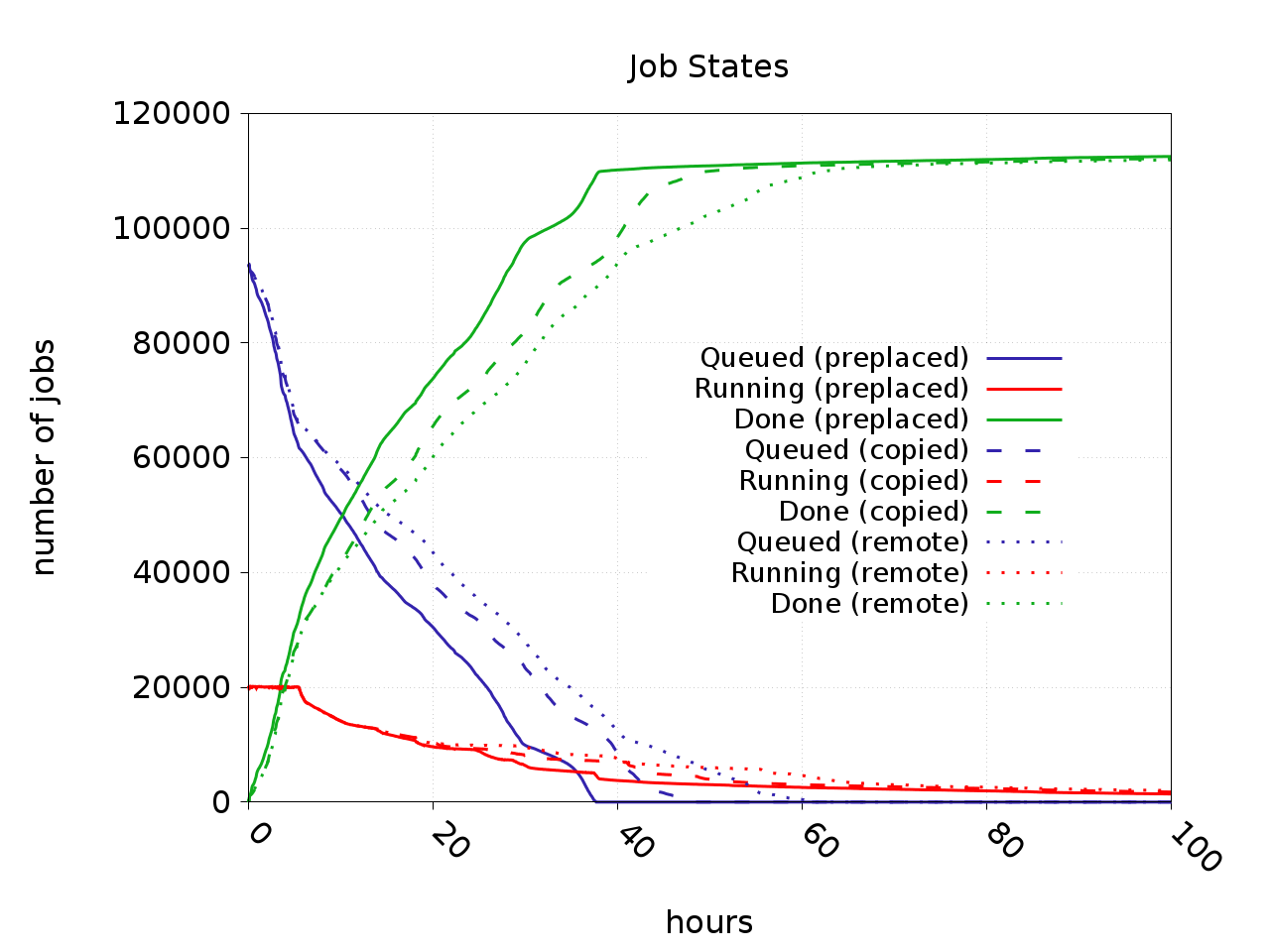}
    \caption{Half CPU/Normal Tran}
  \end{subfigure}
  \begin{subfigure}{0.3\textwidth}
    \includegraphics[width=\textwidth]{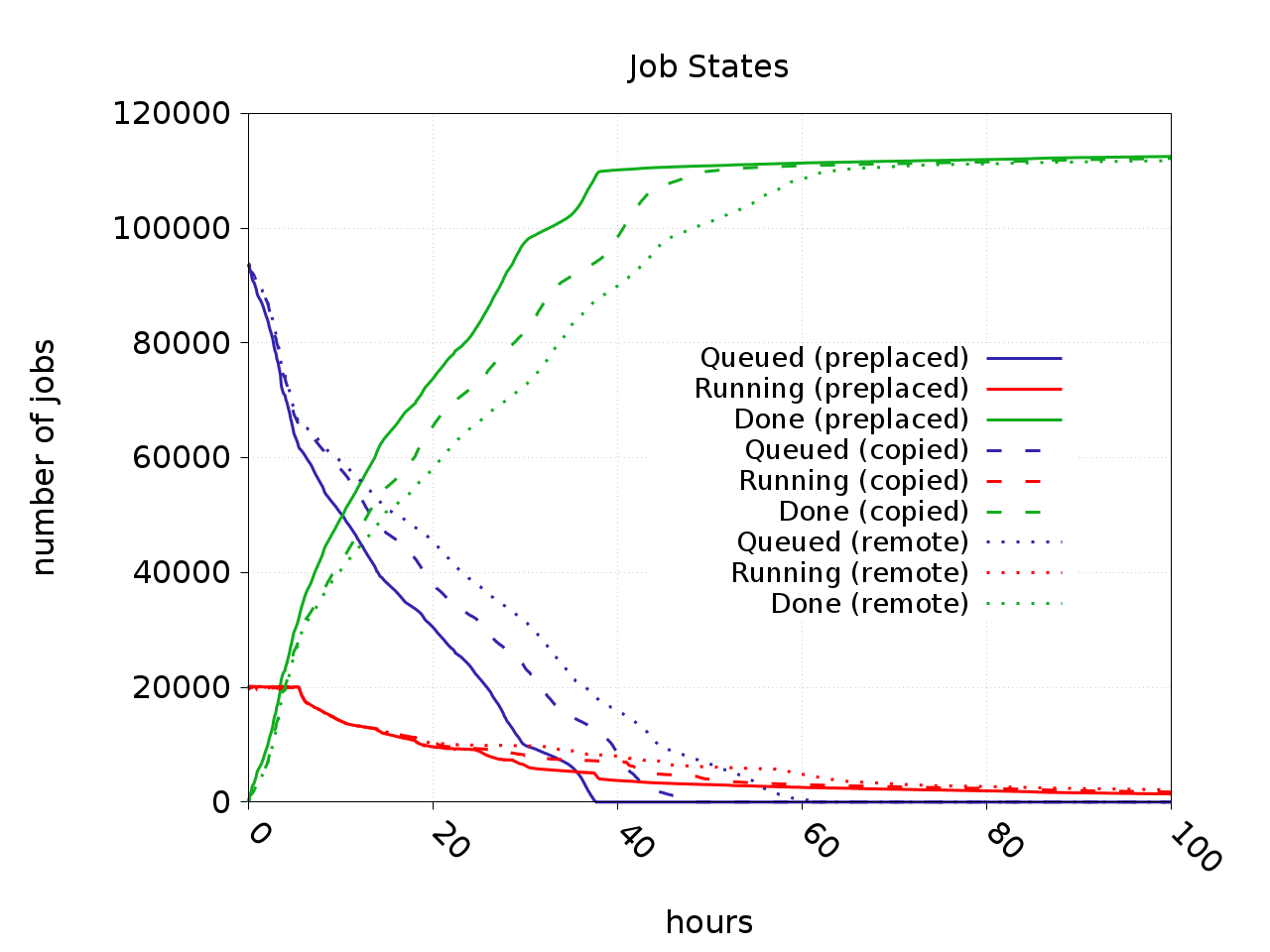}
    \caption{Normal CPU/Normal Tran}
  \end{subfigure}
  \begin{subfigure}{0.3\textwidth}
    \includegraphics[width=\textwidth]{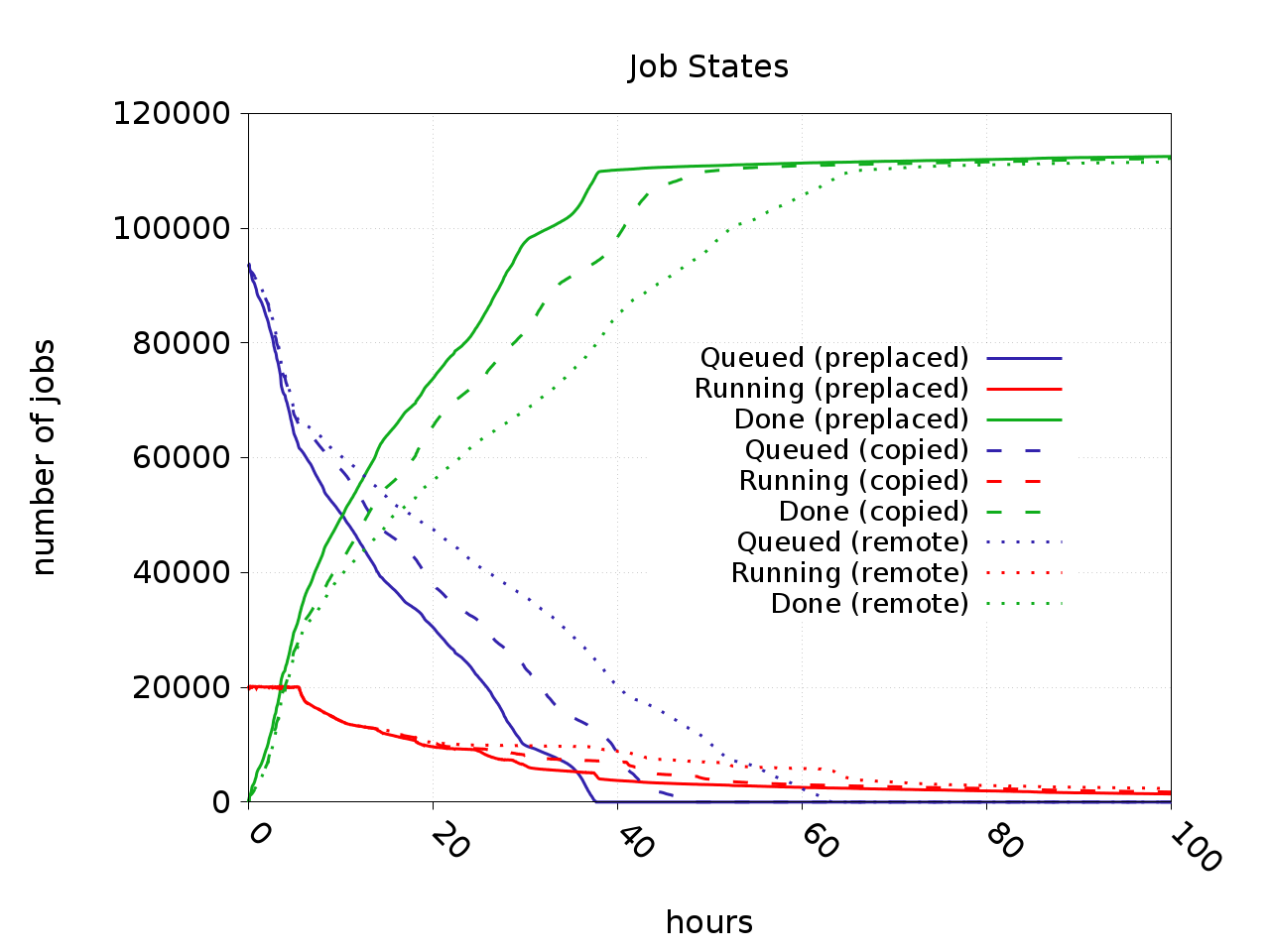}
    \caption{Double CPU/Normal Tran}
  \end{subfigure}
  \begin{subfigure}{0.3\textwidth}
    \includegraphics[width=\textwidth]{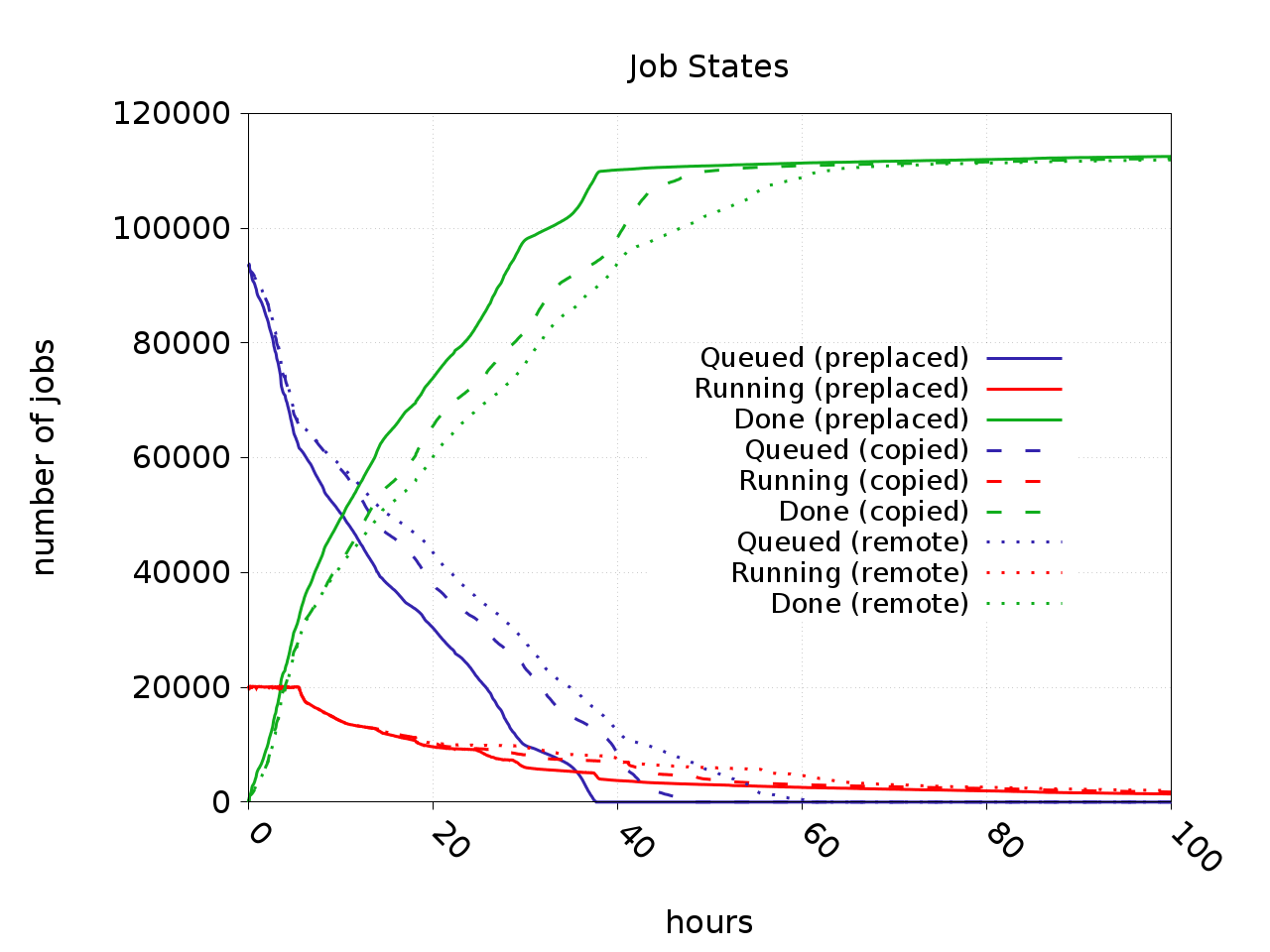}
    \caption{Half CPU/Double Tran}
  \end{subfigure}
  \begin{subfigure}{0.3\textwidth}
    \includegraphics[width=\textwidth]{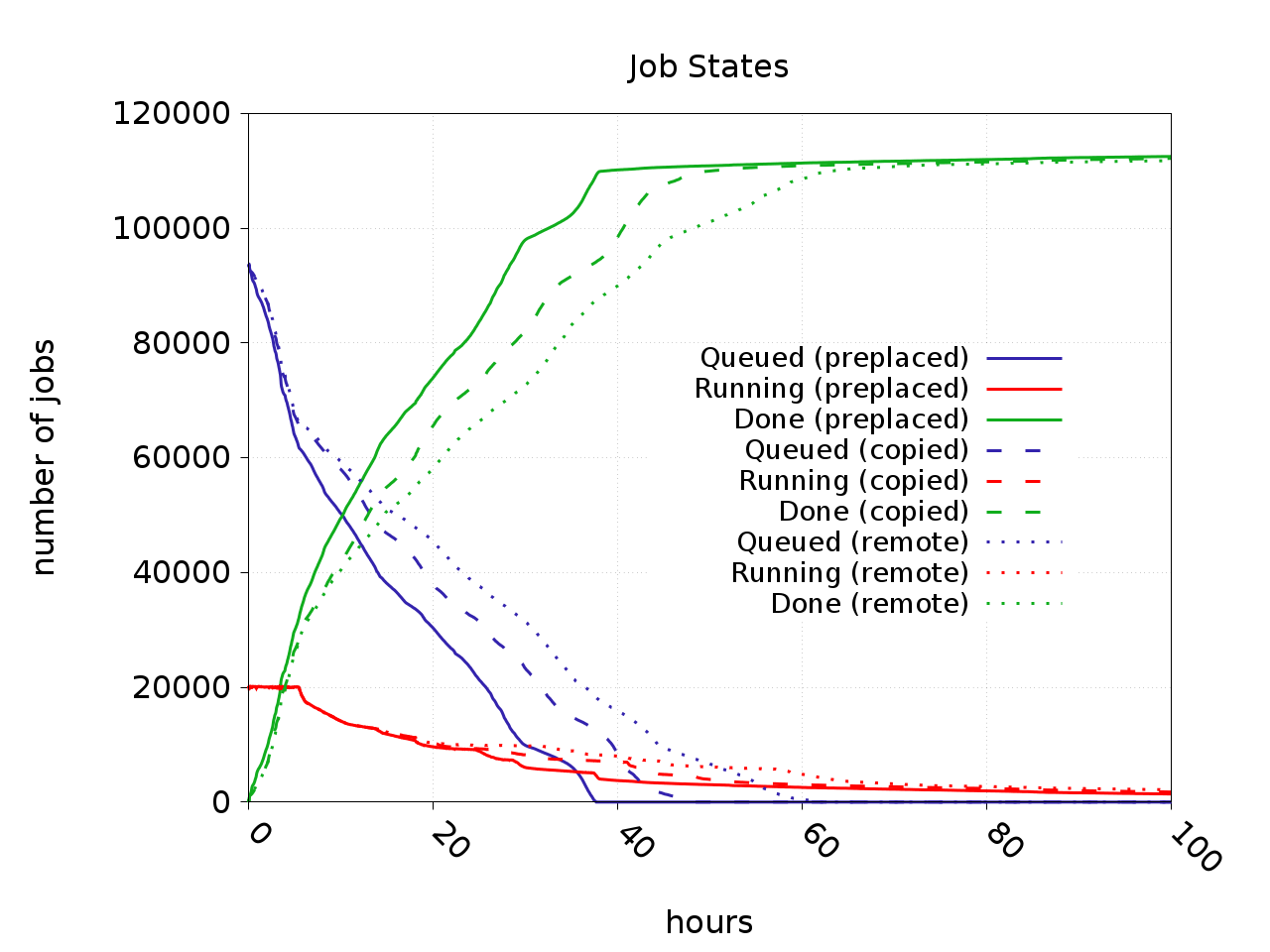}
    \caption{Normal CPU/Double Tran}
  \end{subfigure}
  \begin{subfigure}{0.3\textwidth}
    \includegraphics[width=\textwidth]{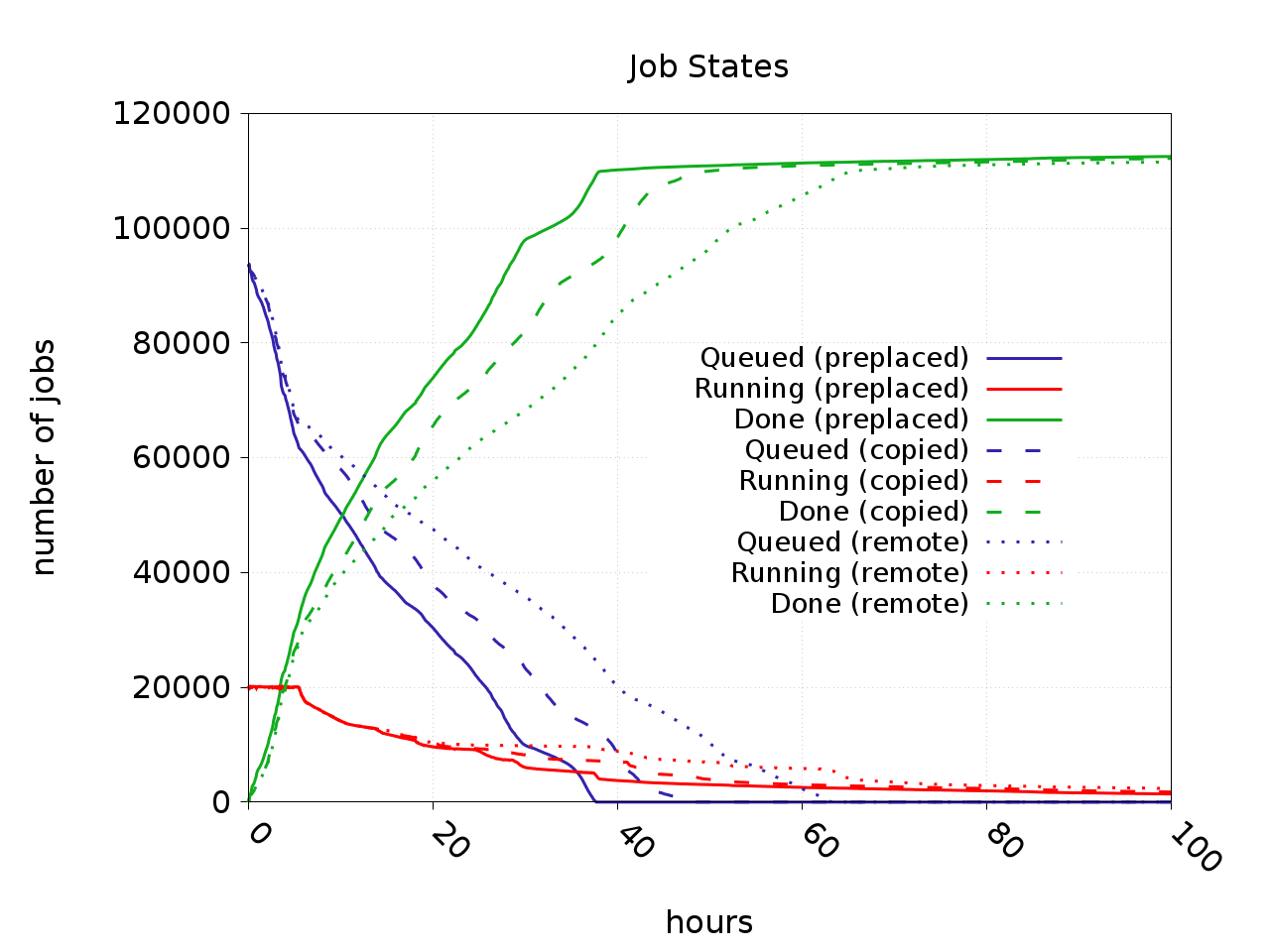}
    \caption{Double CPU/Double Tran}
  \end{subfigure}
  \caption{Sum of job queues at all sites\label{fig:jobQueues}}
\end{figure}

While the simulation is running the state of each of the job lists at
the sites is monitored. We can graph these to see the progression of
jobs through the system. Figure \ref{fig:jobQueues} shows these queues
for each of the nine sets of input parameters.

\subsection{Network Usage}

Another observation from the simulation is the network bandwidth used
in each case. Each of the scenarios has been studied with varying the
input parameters. The variation is in line with expectations and only
the ``normal'' parameter set will be shown for each of the scenarios.

\begin{figure}
  \centering
  \begin{subfigure}{0.3\textwidth}
    \includegraphics[width=\textwidth]{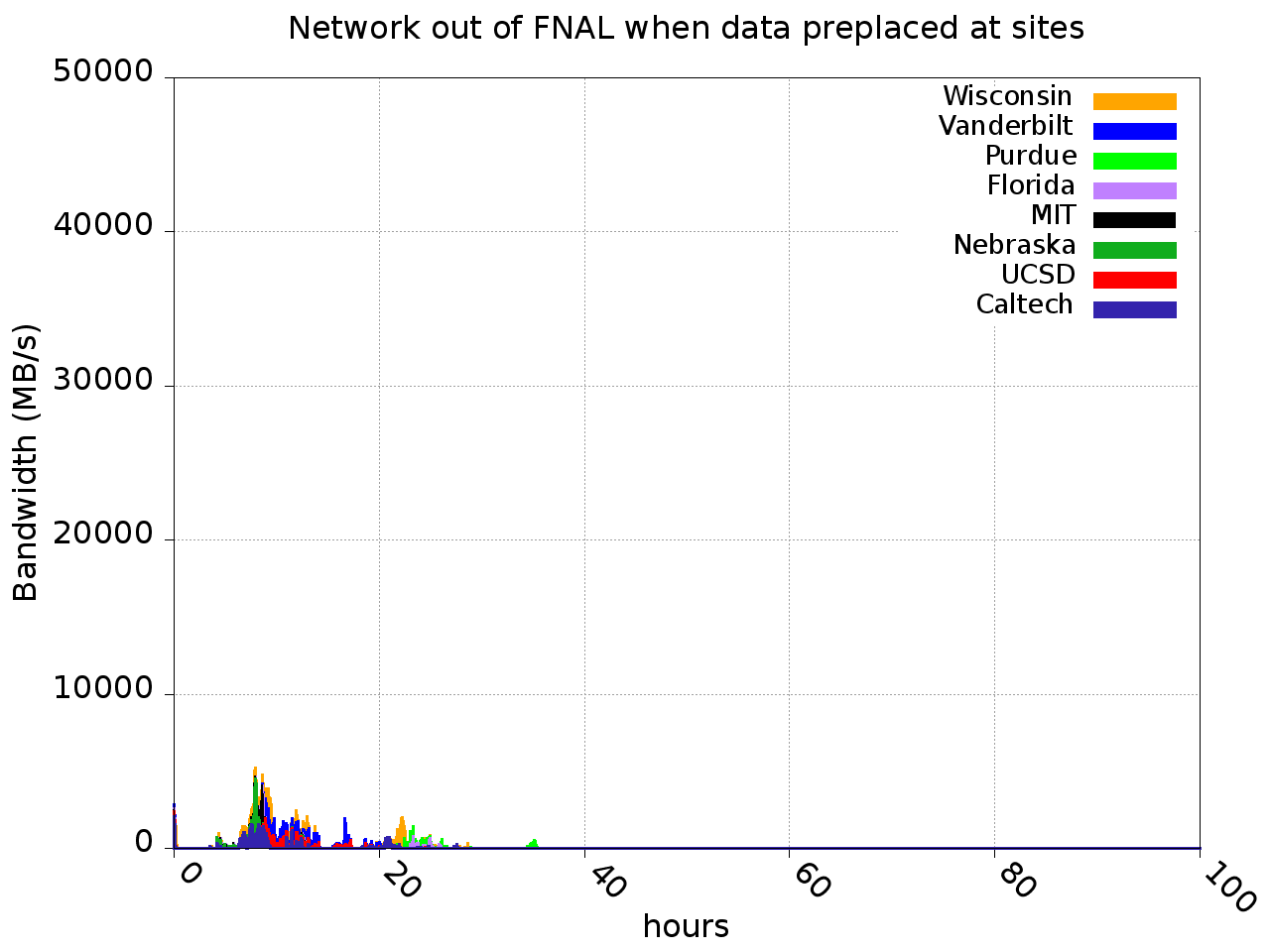}
    \caption{Preplaced Data\label{fig:fnalOutToday}}
  \end{subfigure}
  \begin{subfigure}{0.3\textwidth}
    \includegraphics[width=\textwidth]{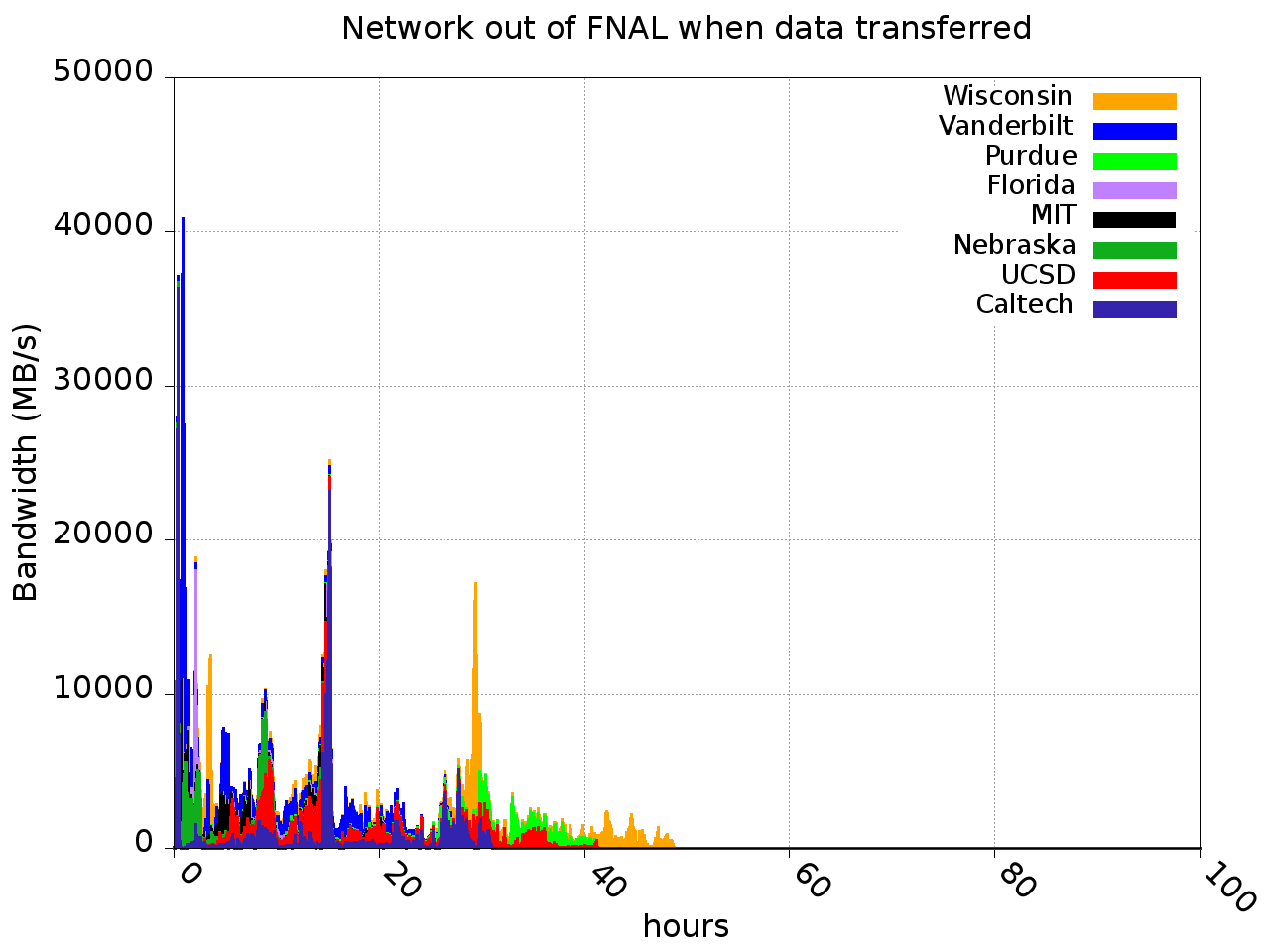}
    \caption{Data Copied\label{fig:fnalOutCopy}}
  \end{subfigure}
  \begin{subfigure}{0.3\textwidth}
    \includegraphics[width=\textwidth]{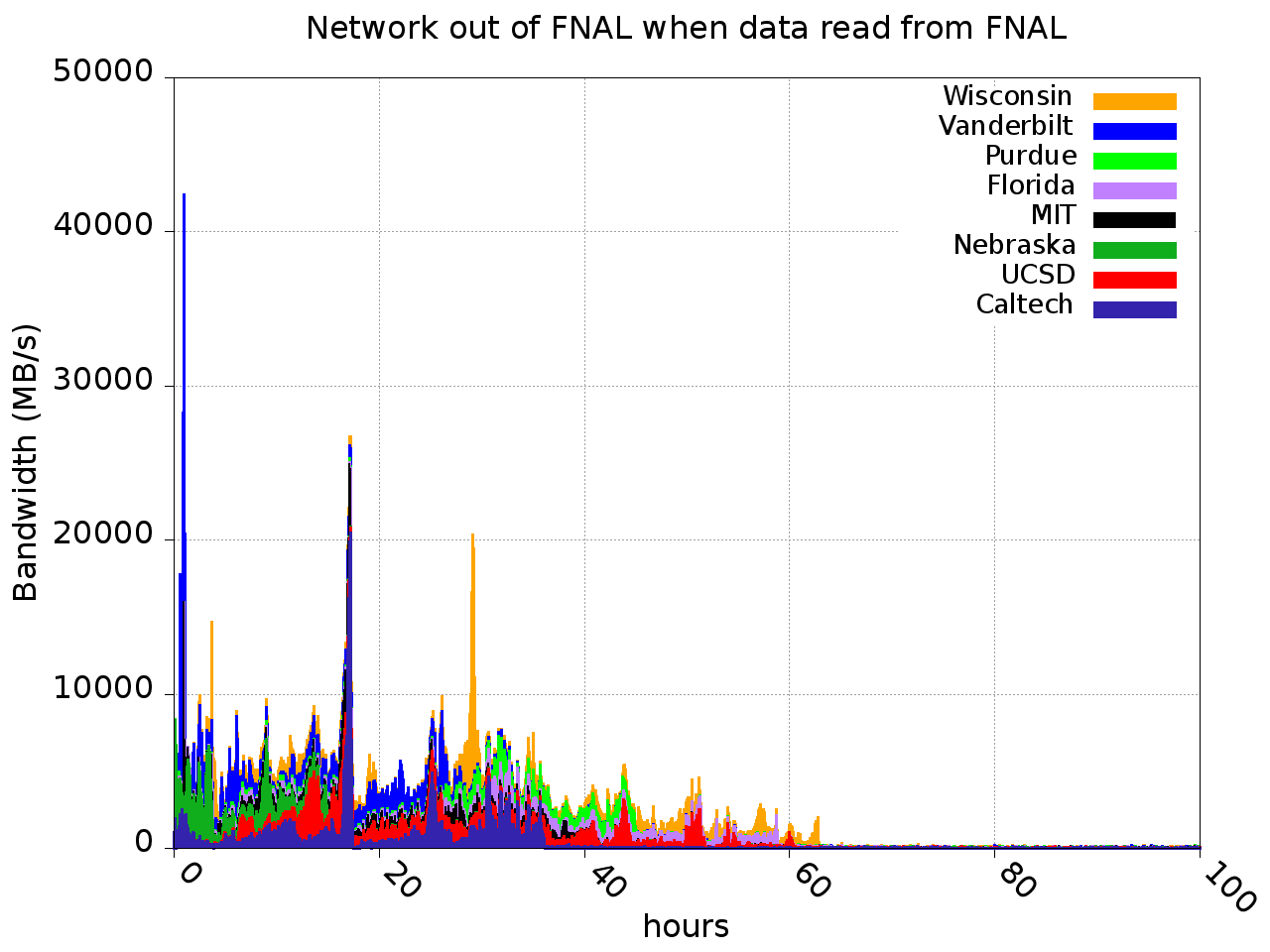}
    \caption{Remote Read\label{fig:fnalOutRemote}}
  \end{subfigure}
  \caption{Bandwidth used out of FNAL\label{fig:fnalOut}}
\end{figure}

In Figure \ref{fig:fnalOut} you are able to see the network out of
FNAL to each of the Tier-2 sites. As expected when data is preplaced
(Figure \ref{fig:fnalOutToday}) at the Tier-2 sites the network used
is very minimal. If data is copied (Figure \ref{fig:fnalOutCopy}) the
network usage is more uneven as the transfers happen as quickly as
possible. Whereas when data is read remotely (Figure
\ref{fig:fnalOutRemote}) data is only consumed at the speed of the job
on the Tier-2 site.

\begin{figure}
  \centering
  \begin{subfigure}{0.3\textwidth}
    \includegraphics[width=\textwidth]{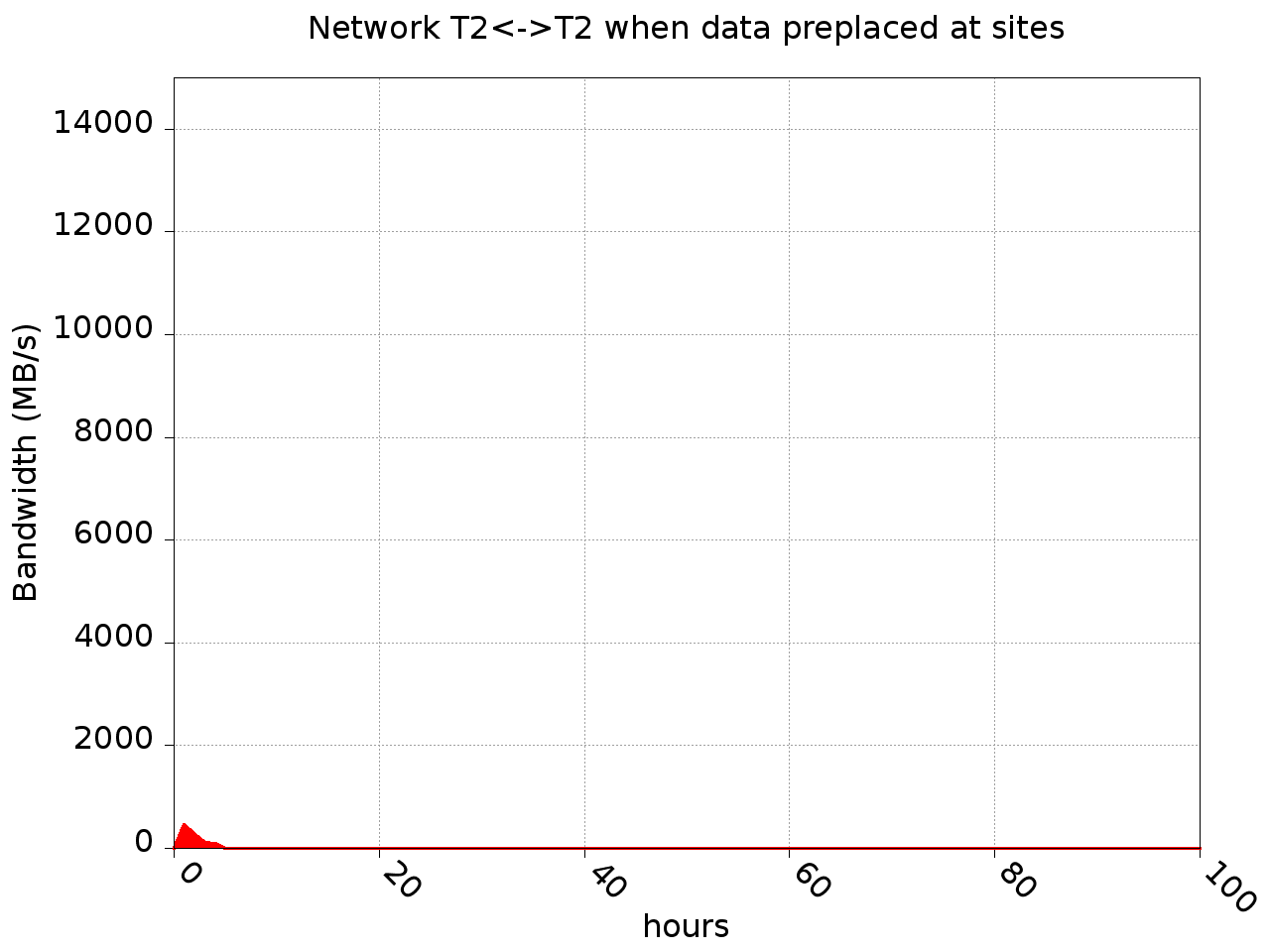}
    \caption{Preplaced Data\label{fig:tier2Today}}
  \end{subfigure}
  \begin{subfigure}{0.3\textwidth}
    \includegraphics[width=\textwidth]{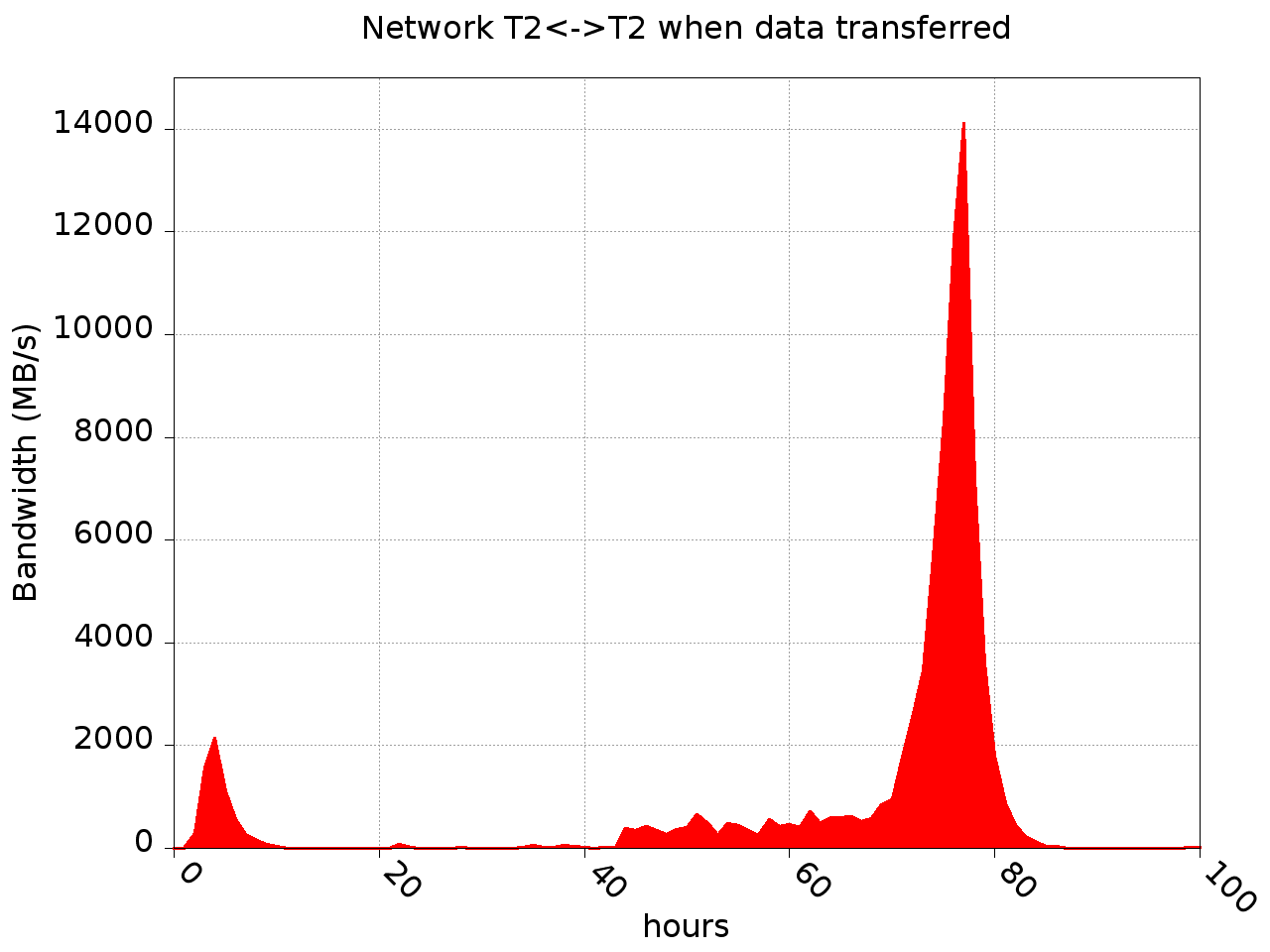}
    \caption{Data Copied\label{fig:tier2Copy}}
  \end{subfigure}
  \begin{subfigure}{0.3\textwidth}
    \includegraphics[width=\textwidth]{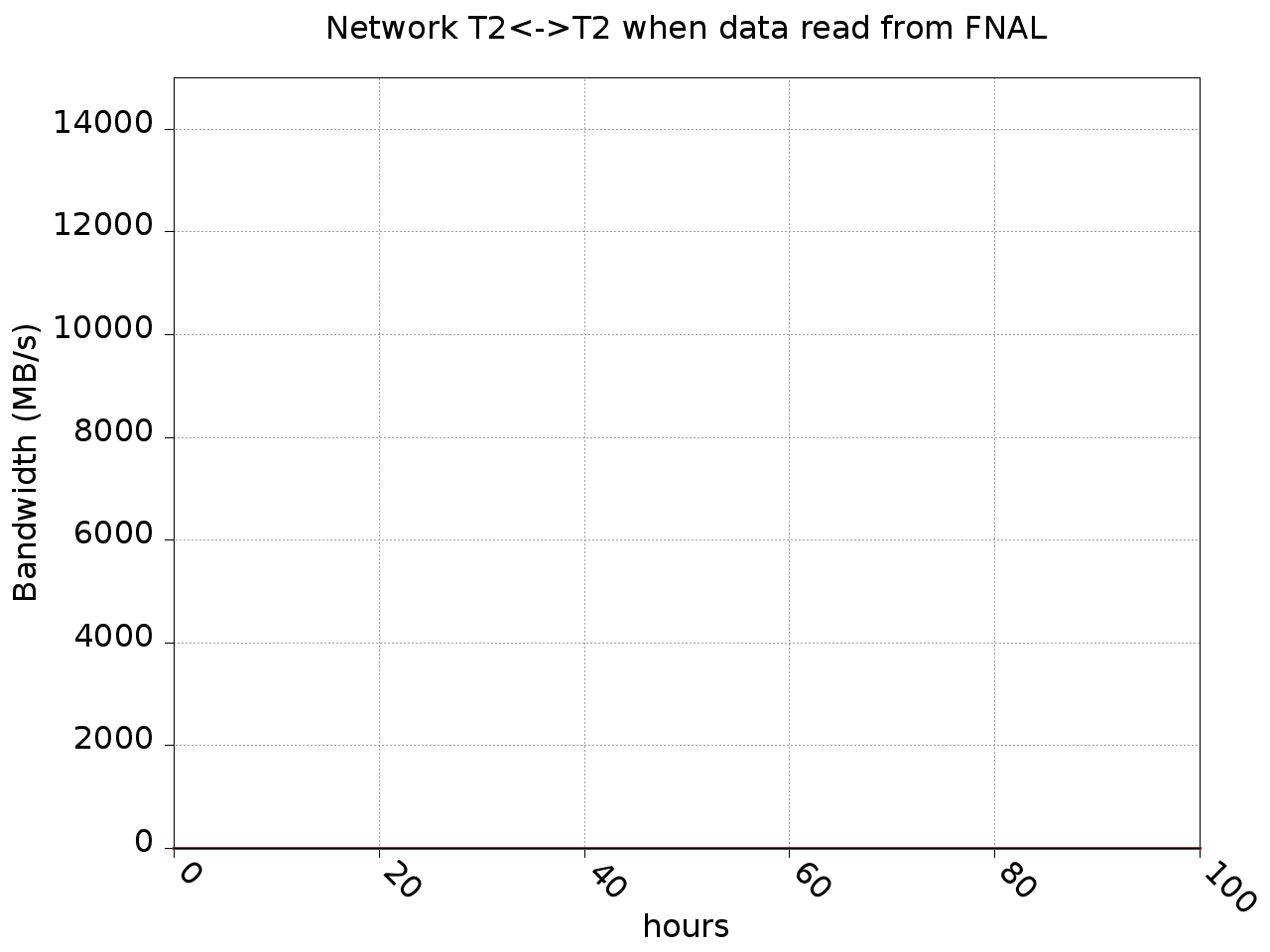}
    \caption{Remote Read\label{fig:tier2Remote}}
  \end{subfigure}
  \caption{Bandwidth used between Tier-2 sites\label{fig:tier2}}
\end{figure}

The network load between the Tier-2 sites was also examined. The total
Tier-2 to Tier-2 bandwidth used is show in figure \ref{fig:tier2},
again only the ``normal'' parameter set is shown. For the case when
the majority of the data is preplaced (Figure \ref{fig:tier2Today}) at
the job execution site there is very little Tier-2 to Tier-2
activity. The small amount of data transferred will be for those jobs
which ran at a site which didn't have the data locally but was
available from another Tier-2 which was ``nearer'' than FNAL. For the
case when the data was copied from FNAL (Figure \ref{fig:tier2Copy})
the larger proportion at the end will be data duplicated from FNAL to
the Tier-2 sites, which then represented a better replica that the
original copy at FNAL for a later job ran at another Tier-2 site. For
the case when data is always read from FNAL (Figure
\ref{fig:tier2Remote}) there are no replicas made at Tier-2s, so there
is no Tier-2 to Tier-2 data transfers.

\section{Conclusions}

The simulation can show us what can happen in different scenarios. As
is shown with the situations studied in this paper alternative
strategies don't produce results quicker, however, there could be
significant savings in cost and maintenance of disk space at Tier-2
sites, which could allow purchase of more CPU resources to offset the
extra wall clock time required. Once a more accurate set of simulation
parameters has been determines quantitative comparisons can be made.

\end{document}